\documentclass{aa}  
\usepackage{graphicx}
\usepackage{txfonts}
\usepackage{physabbrv}
\usepackage[colorlinks=true,citecolor=blue,linkcolor=blue]{hyperref}

\usepackage{caption}
\usepackage{subcaption}

\usepackage{xcolor}
\usepackage{amsmath}

\newcommand{\Teff}{T_{\rm{eff}}}

\defcitealias{Okamoto2021}{O2021}

\newcommand{\NLC}{2274 }
\newcommand{\NF}{342 }
\newcommand{\NFTS}{283 }
\newcommand{\NSTS}{178 }
\newcommand{\NFBS}{47 }
\newcommand{\NSBS}{29 }

\newcommand{\NFBOTS}{10 }
\newcommand{\NSBOTS}{4 }
\begin{document} 

\let\oldpageref\pageref
\renewcommand{\pageref}{\oldpageref*}

%   \title{Super-flares on solar-like stars}
%   \subtitle{I. A new method for identifying the true flare locations on the Kepler CCDs}

    \title{Superflares on solar-like stars:}
    \subtitle{A new method for identifying the true flare sources in photometric surveys}

    \author{V. Vasilyev\inst{1} 
        \and T. Reinhold\inst{1}
        \and A. I. Shapiro\inst{1} 
        \and N. A. Krivova\inst{1} 
        \and I. Usoskin\inst{2}
        \and B. T. Montet\inst{3,4}
        \and S. K. Solanki\inst{1}
        \and L.~Gizon\inst{1, 5, 6}
        }

   \institute{Max-Planck-Institut f\"ur Sonnensystemforschung, Justus-von-Liebig-Weg 3,   
            37077 G\"ottingen, Germany \\
              \email{vasilyev@mps.mpg.de}
        \and
            Sodankyl\"a Geophysical Observatory and Space Physics and Astronomy Research unit, FIN-90014, University of Oulu, Finland  
        \and 
        School of Physics, University of New South Wales, Sydney, NSW 2052, Australia
        \and 
        UNSW Data Science Hub, University of New South Wales, Sydney, NSW, 2052, Australia
        % \and 
        %     School of Space Research, Kyung Hee University, Yongin, Gyeonggi 446-701, Korea
        \and 
            Institut für Astrophysik, Georg-August-Universit\"at G\"ottingen,  G\"ottingen, Germany
        \and  
            Center for Space Science, NYUAD Institute, New York University Abu Dhabi, Abu Dhabi, UAE  
            }

   \date{Received ...; accepted ...}

% \abstract{}{}{}{}{} 
% 5 {} token are mandatory
 
  \abstract
  % context heading (optional)
   {Over the past years, thousands of stellar flares have been detected by harvesting data from large photometric surveys. These detections, however, do not account for potential sources of contamination such as background stars or small solar system objects appearing in the same aperture as the primary target.}
  % aims heading (mandatory)
  {We present a new method for identifying the true flare sources in large photometric surveys using data from the Kepler mission as an illustrative example. The new method considers not only the brightness excess in the stellar light curves but also the location of this excess in the pixel-level data.}
  %The method is based on fitting the point spread function in the pixel-level data with the aim of associating the flux excess in the image to the target star or some other source of contamination.]}
  % methods heading (mandatory)
  {Potential flares are identified in two steps: First, we search the light curves for at least two subsequent data points exceeding a $5\sigma$ threshold above the running mean. For these two cadences, we subtract the "quiet" stellar flux from the Kepler pixel data to obtain new images where the potential flare is the main light source. In the second step, we use a Bayesian approach to fit the point spread function of the instrument to determine the most likely location of the flux excess on the detector. We match this location with the position of the primary target
  and other stars from the Gaia DR2 catalog within a radius of 10 arcsec around the primary Kepler target. When the location of the flux excess and the target star coincide, we associate such events with flares on the target star.}
  % results heading (mandatory)
  {We applied our method to 5862 main-sequence stars with near-solar effective temperatures. From the first step, we found \NLC events exceeding the 5-sigma in at least two consecutive points in the light curves. Applying the second step reduced this number to \NF  superflares. Of these, \NFTS flares happened on \NSTS target stars, \NFBS events are associated with fainter background stars, and in \NFBOTS cases, the flare location cannot be distinguished between the target and a background star. We also present cases where flares have been reported previously but  our technique could not attribute them to the target star.}
  % conclusions heading (optional), leave it empty if necessary 
  {We conclude that 1) identifying outliers in the light curves alone is insufficient to attribute them to stellar flares and 2) flares can only be uniquely attributed to a certain star when the instrument pixel-level data together with the point spread function are taken into account. As a consequence, previous flare statistics are likely contaminated by instrumental effects and unresolved astrophysical sources.}

   \keywords{Stars: flare -- Stars: activity -- Stars: solar-type -- Methods: data analysis}

   \maketitle
%
%-------------------------------------------------------------------

\section{Introduction}
% flares: solar & stellar (Kepler)
Flares are highly energetic outbursts in the stellar atmosphere caused by the reconnection of magnetic field lines. Such events can be detected over a broad wavelength range from radio waves to gamma rays \citep{Benz2010}. On the Sun, flares have been observed in gamma- and X-rays \citep{Ackermann2014Gammarays, Lysenko2020Xrays},  in the UV  and in  white light \citep[see, e.g.,][]{ Haisch1991, Benz2010, Hudson2011, Kretzschmar2011} as well as in the radio band \citep{Bastian1998RadioFlares}. 

Before the era of space telescopes, stellar flares were detected only for a handful of G-type stars \citep{Schaefer2000}. With the advent of large photometric surveys, in particular the \textit{Kepler} mission, it became clear that flares are a common phenomenon on cool stars \citep{Walkowicz2011,Maehara2012,Shibayama2013,Davenport2016,Doorsselaere2017,Yang2019,Guenther2020,Ilin2021}. In particular, it was shown that G-dwarfs can experience very energetic flares (termed superflares) with radiative energies up to $10^{36}$ erg \citep{Maehara2012}. This energy is several orders of magnitude larger than the radiative energies of solar flares, which are usually well below $10^{32}$ erg \citep{Hudson2011}.  The highest solar flare energy ever detected was $\sim4 \times 10^{32}$ erg \citep{Emslie2012}.

Can the Sun also unleash a superflare, and if so, how often does it happen? This question is of crucial importance since superflares occurring on the Sun might lead to devastating consequences for humanity. The solar record alone is not sufficient to answer this question. At present, the only way to understand whether the Sun has experienced superflares is to search for extreme solar energetic particle (SEP) events in the cosmogenic-isotope data \citep[e.g.,][]{usoskin_LR_17,miyake19}. These events are thought to be associated with very energetic flares, and presently, eight extreme SEP events (five are confirmed while three are still pending) are known during the 12 millennia of the Holocene \citep{usoskin_GRL_21, Brehm2021, Brehm2022}. However, the relation between superflares and extreme solar particle events is still highly controversial, given that they are based on different physical manifestations \citep[visible electromagnetic emission vs. fluence of energetic particles, respectively, see, e.g.,][]{usoskin_LR_17}.

Another way to understand whether the Sun is capable of unleashing superflares is provided by the solar-stellar comparison. Studies of stellar superflares have been progressively focusing on stars similar to the Sun, i.e., stars with near-solar effective temperatures and rotation periods. Constraining these two parameters is important because they define the action of stellar dynamos, and, consequently stellar magnetic activity \citep[see, e.g., the review by][]{Ansgar2012}. 

In one of the first studies based on \textit{Kepler} data, \citet{Maehara2012} found 14 superflares on stars with near-solar effective temperatures $T_\mathrm{eff}=5600-6000$\,K and rotation periods, $P_\mathrm{rot}$, longer than 10 days (for comparison, the solar sidereal Carrington rotation period is 25.4 days). \citet{Maehara2012} concluded that these stars unleash superflares with energies larger than $10^{34}$ erg (X1000-class flare in solar terminology)  once every 800 years. However, most of the flaring stars studied by \citet{Maehara2012} rotate significantly faster than the Sun, so their results are not directly applicable to the Sun.  Consequently, more recent studies focused on slower rotators. For example, \citet{Notsu2019} used measurements of stellar rotational periods by \cite{McQuillan2014} to conclude that stars with near-solar $T_\mathrm{eff}$ and $P_\mathrm{rot}\sim 25$ days should unleash superflares every 2000--3000 years. A more recent analysis by \citet{Okamoto2021} concluded that the Sun can unleash  X1000-class superflares once every 6000 years.

Crucially, all previous estimates of flare occurrence rates on solar-like stars have been based on the analysis of stars with {\it known rotation periods}. At the same time it has been shown that, if the Sun were observed by {\it Kepler}, it would most probably be classified as an inactive star with an unknown rotation period \citep{Aigrain2015, Reinhold2020, Shapiro2020, Amazo2020, Reinhold2021}. As a result, studies focused on stars with known rotation periods  exclude most of the stars truly similar to the Sun which might bias the statistics of superflares toward more active stars \citep[][]{Reinhold2020}.

In this context, we take a different data-based approach for correcting the bias introduced by stars with unknown rotation periods. Namely, we consider a larger sample of stars, mostly following the approach of \citet{Reinhold2020}.
These authors studied the variability of main-sequence stars with near-solar effective temperatures and metallicities. Thereof, the majority of stars had unknown rotation periods but showed photometric variabilities very similar to the Sun. This sample of solar-like stars will be analyzed for flares here.

% method & goal
When comparing solar and stellar flare statistics, it is important to keep in mind that highly energetic events are extremely rare on truly solar-type stars. Thus, every single flare detection will have a big impact on the statistics. This calls for accurate and stable methods of flare identification in large stellar samples. 

In stellar light curves, flares are seen as sudden brightness increases, followed by an exponential decay that typically lasts for several hours. Thus, previous flare studies were based on searching for outliers in the photometric time series. This approach, however, does not account for the possibility of flares occurring on background stars or other sources of brightness variations in the telescope's field of view. Here, we present a new method that is based on a combined analysis of the { \it Kepler} stellar light curves, {\it Kepler} target pixel files, and the high-precision astrometric data from the Gaia mission. This combination of data allows us to determine the most probable location of the brightness increase on the detector, and thus to distinguish between flares on the target star and  background stars unresolved by the survey telescope. This paper aims to present this new approach and apply it to a large sample of solar-like stars to gain better statistics on their flare frequencies. A statistical analysis of the results and its implication for the Sun will be the subject of a follow-up paper.

\section{Data}\label{sect:data}
% describe TPF
In our analysis, we use two different data products of the Kepler telescope: the target pixel file (TPF) and the resulting light curves. The TPF contains a series of images centered at the target star. The size (in pixels) of each image depends on the target star (with an image scale of 3.98 arcsecs per pixel). 

Owing to the optical system of the telescope, the target flux is distributed over several pixels. To extract the flux of the target star, different aperture masks have been tested to sum up the flux of individual pixels. Before this step, the TPF underwent procedures of cosmic-ray removal and background corrections to maximize the flux of the target star. For each exposure, the sum of the pixels in the aperture equals the so-called Simple Aperture Photometry (SAP) flux.

The Kepler light curves are one-dimensional time series with flux measurements of the target stars. These have been subject to various corrections for instrumental systematics \citep{Stumpe2012, Smith2012, Stumpe2014}. In this study, we use the Pre-search Data Conditioning (PDC-SAP) flux from the latest data release (DR25), for which most instrumental effects have been removed. We only use the Kepler long-cadence (LC) data products\footnote{Kepler data products are available at \url{https://archive.stsci.edu/missions-and-data/kepler}} with a cadence of $\Delta t \approx 29.4$ minutes. Kepler data were released in 18 quarters (Q0-Q17), with most of them covering $\sim90$ days each (with the exception of Q0, Q1, and Q17, which span a shorter time). We analyze all available light curves and target pixel files (TPF) from quarters 0--17 for the entire observing period from May 2009 to May 2013.

The target list that we will test our method on is based on the sample of stars with near-solar effective temperatures studied by \citet{Reinhold2020}. These authors selected Kepler stars brighter than 15th magnitude with effective temperatures between $5500 < \Teff < 6000$\,K, surface gravities $\log\,g>4.2$, and metallicities in the range $-2.0 < \rm{[Fe/H]} < 0.5$ dex, using the catalog of \citet{Mathur2017}. Additionally, we use the Gaia-DR2 catalog\footnote{Data available at \url{https://vizier.u-strasbg.fr/viz-bin/VizieR-3?-source=I/345/gaia2}} \citep{2018yCat.1345....0G} to retrieve astrometric data such as target positions (right ascension and declination), and G-band magnitudes of all stars within a radius of 10 arcsec around each Kepler target. Furthermore, we computed Gaia absolute magnitudes to select only main-sequence stars confined between two isochrones in the Hertzsprung-Russell diagram (for details, see Fig.~1 in \citealt{Reinhold2020}). By cross-matching the selected targets in the Kepler and Gaia archives, we found 5862 Kepler stars matching the selection criteria.

%%%%%%%%%%%%%%%%%%%%%%%%%%%%%%%%%%%%%%%%%%%%%%%%%%%%%%%%%%%%%%%%%%%%%%%%%%%%%%%%%%%%%%%%%
\section{Methods}
%%% plot this example result_kic3936679_q12_t1143.136.png
\label{sec:method}
To identify flares on the selected stars, we employ a two-step algorithm. First, we collect potential flare events by searching the stellar light curves for data points exceeding a certain threshold. This approach is similar to the methods applied in earlier studies, although with some modifications (Sect.~\ref{sec:lightcurves}).
%We also classify the flare candidates into several groups depending on the patterns of the brightness changes during the potential flare events.
Second, we localize the position of each event on the detector by fitting the instrument's point spread function (PSF).
% to the residual image, computed by removing the quiet stellar light from the TPF (Sect.~\ref{sec: flare_localization}). Thus, the residual image contains a flux excess due to the flare. 
If the positions of the target star and the event coincide, we conclude that the flare is associated with the target star (Sect.~\ref{sec: flare_localization}). These two steps are described in detail below.

%\section{Contamination}
\subsection{Searching for flares in the light curves}
\label{sec:lightcurves}
\begin{figure*}[ht]
\centering
    \includegraphics[width=1.0\textwidth]{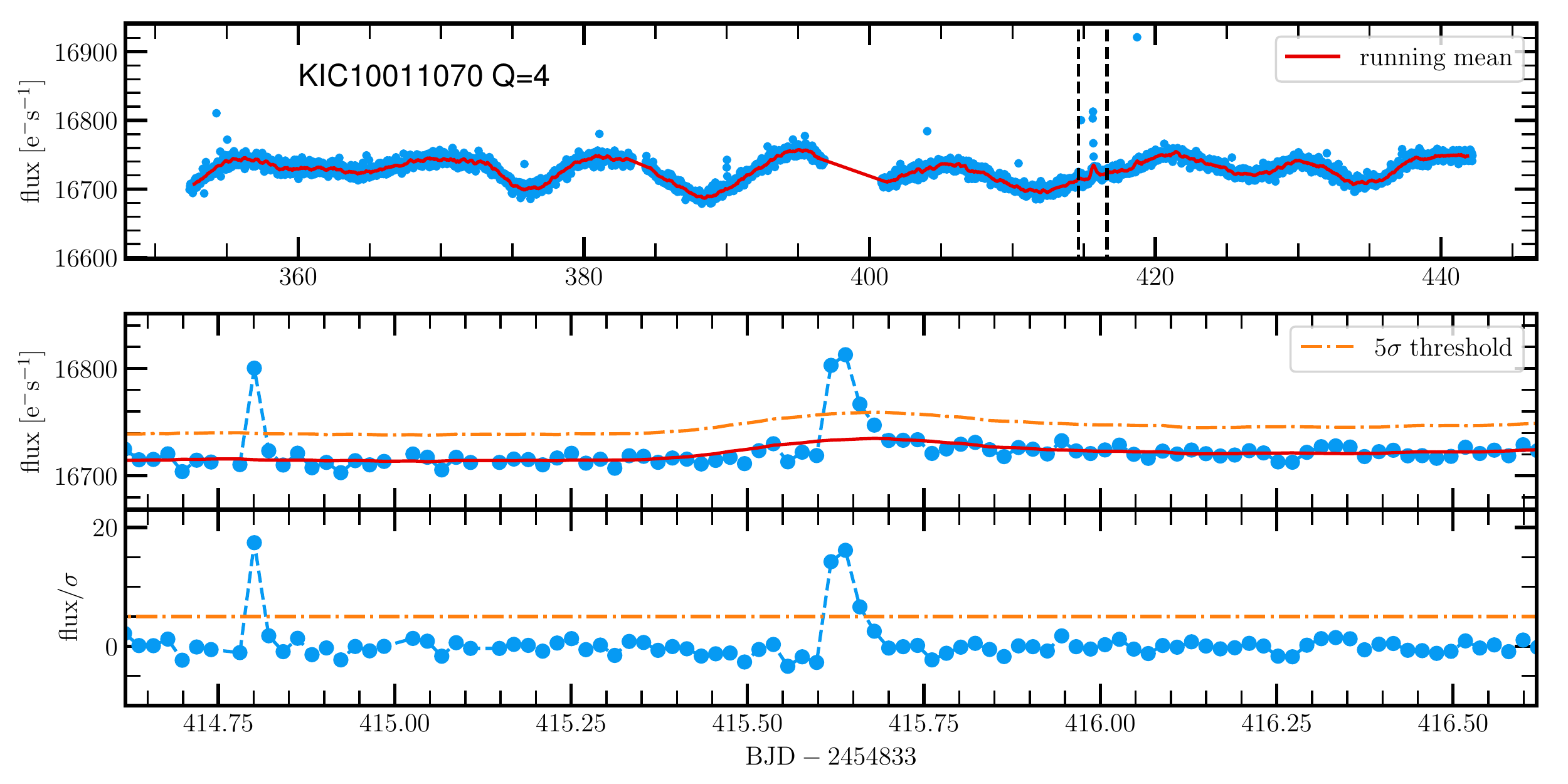}
    \caption{Detection of a flare candidate in the light curve of the star KIC\,10011070 observed in quarter $Q=4$.
    \textit{Top panel}: Original light curve (blue dots) and running mean (red line) calculated with a boxcar function averaging over $15$ cadences. The vertical dashed lines mark a period around a potential flare.
    \textit{Middle panel}: Zoom-in to the period around the flare candidate, as indicated in the top panel. The running mean is shown as solid red line and the dashed-dotted orange line shows the running $5\sigma$ threshold.
    \textit{Bottom panel}: Detrended time series by subtracting the running mean and normalizing by the standard deviation $\sigma$. The three consecutive data points in the middle clearly exceed the $5\sigma$ threshold.}
    \label{fig:lc_flare_search_in_light_curve}
\end{figure*}

Our first goal is to find potential flare events in the light curves of the target stars and identify their time of occurrence. 
Several flare detection methods have been reported \citep[e.g.,][]{Shibayama2013, Johnson2017, Ilin2021}. Generally, these methods are based on removing any variability on rotational and longer time scales from the light curves and searching for outliers in the detrended time series. More recent methods often use a Bayesian approach \citep{Pitkin2014, Guenther2020} or machine-learning \citep{Vida2018, Feinstein2020}.

Here, we use a simple moving average filter with the width of $M = 15$ data points (i.e. $\sim 7.5$\,h) to detrend the time series. This approach reaches its limits when the width of the window becomes comparable to the stellar rotation period. In this study, however, we analyze solar-like stars that either have rotation periods $>20$ days, as measured previously \citep{McQuillan2014}, or do not have periods reported yet. The latter makes it unlikely that the target star has a very short period ($\lesssim 1$ day) because short periods are detected more easily, and would likely have been measured before. The choice of the optimal flare detection filter is beyond the scope of this study.

The running mean filter is sensitive to discontinuities that are often found in the Kepler light curves. To reduce their impact, we excluded $M$ cadences on either side of all data gaps longer than $\Delta t M/4\approx 2$ hours. Next, we applied the running mean filter, subtracted the smoothed light curve from the original one, and obtained a detrended time series. We excluded the first and the last $M$ cadences from the detrended light curve due to the edge effects of the running mean. The fraction of rejected data points is around 3.2\%. However, we noticed that the detrended light curves contain bumps due to individual outliers in the original light curve. Their impact can be reduced by either excluding or replacing them in the original light curves before applying the running mean filter. Firstly, we excluded 1\% of points with the lowest and highest flux in the detrended time series and computed the standard deviation $\sigma$ for the remaining time series. Secondly, we found outliers that are above the chosen threshold of $2.5\sigma$, replaced them in the non-detrended light curves, and repeated the procedure described above.

We identified potential flare events in the detrended light curves using the following criteria: (i) the event consists of at least two consecutive data points above the threshold set to $5\sigma$, and (ii) there are no data gaps within 1.5 hours before and after these points. The first criterion is set to detect the most energetic flares with high statistical confidence. The second criterion reduces the risk that some outliers might be related to instrumental effects, e.g., an increased noise level around the gap and statistical uncertainties caused by missing data. The moment in time of the first data point above $5\sigma$, we define as the time when the flare starts and denote it as $t_\mathrm{flare}$. Typically, after reaching the maximum, the flare flux exponentially decays and, in the end, the light curve reaches a similar level as before the flare. In the descending phase, we define the endpoint of the flare as the time when the flare flux crosses the $1\sigma$ level again. This point is defined by linear interpolation between the last data point above and the following one below the $1\sigma$ level. Consequently, the interval between the flare start and endpoints is defined as the flare duration, denoted as $\delta t_\mathrm{flare}$.

Figure~\ref{fig:lc_flare_search_in_light_curve} shows an example of the flare search in the light curve for the star KIC\,10011070. We found a potential flare at $t_\mathrm{flare}=415.62$ days satisfying our criteria described above.

\subsection{Localization of flares in the target pixel files}\label{sec: flare_localization}
After identifying all potential flare candidates in the light curves, we use the series of images around each potential event (i.e. the target pixel files) to localize the source of the brightness excess in the sky. As described in Sect.~\ref{sect:data}, the Kepler pipeline generates the light curves using photometric aperture masks. These masks define the pixels on the CCD over which the flux of a given target is summed for each cadence. Potentially, a (much) fainter background star (below the sensitivity of Kepler) may fall in the same aperture as the target star. A flare on such a background star might be detectable in the final light curve but would be misinterpreted as a flare on the target star. To address this, we localize the observed flux excess on the detector and match this location with the positions of the target and potential background stars from the Gaia DR2 catalog. 

When starlight passes the optical system of a telescope, it is spread over the detector by the Point Spread Function (PSF). Often, the PSF is wider than the pixel size of the detector, and therefore the starlight is distributed over several pixels. The image of a target recorded on the detector, known as the Pixel Response Function (PRF), is a combination of the true image of the object produced by the optical system and the detector sensitivity. Thus, a reconstruction of the position of a star on the image requires knowledge of the shape of the PRF for a given instrument. The PRF of the Kepler instrument is a combination of the optical PSF, the spacecraft pointing jitter, the module defocus, the CCD response function, and the electronic impulse response (for details, see \citealt{Kepler_PRF_Bryson} and \citealt{Kepler_Instrument_Handbook}). Extraction of the photometry by fitting the PSF/PRF is called PSF-fitting.

The flare localization is based on the following key points. For each potential flare event, we select a series of images (target pixel files) before and after $t_\mathrm{flare}$. These time intervals represent the quiet stellar flux in the absence of the flare. The flare creates an additional flux excess on top of the quiet stellar flux in the pixels close to the center of the target star during $t_\mathrm{flare}$ and the following cadences. We aim to remove the quiet stellar flux to obtain a new series of images, where the flare is the main light source. Then, we model the flare at each cadence as a point source with a given position and flux convolved with the Kepler PRF and use a Bayesian approach to find the most likely flare location. We describe each of these steps in detail below.

Let us denote the image of the target star taken at the time $t_k$ as $ F (t_k) =\{ F_{p}(t_k)\}$, where $k$ is the cadence,  $p$ is the pixel index, and $F_{p}(t_k)$ is flux $[\mathrm{e}^{-}\mathrm{s}^{-1}]$ in the pixel $p$. Let $N_x$ and $N_y$ be the number of columns and rows in the image, where $x$ (column number) and $y$ (raw number) are the coordinates of a pixel on the two-dimensional image. Let $N_\mathrm{pix}=N_x N_y $ be the total number of pixels in the image and $1 \leq p \leq  N_\mathrm{pix}$. For a given flare, we select images within a time interval $\Delta T=0.6$ days centered at $t_\mathrm{flare}$ for this analysis.

As described above, the image of the target star might also contain contaminant stars. Let us denote the total stellar flux of all stars in a given pixel $p$ during the absence of flares (i.e. `quiet` stellar flux) as $S^\star_p(t_k)$. A flare with the flux $f(t_k)$ and coordinates $(x(t_k), y(t_k))$ at a certain cadence $t_k$ introduces an additional flux in the image pixels because its light is distributed by the PSF. The flux excess from the flare in each pixel $\Delta F_p(t_k)$ is given by the following expression: 
\begin{equation} \label{eq:flare_model}
\begin{aligned}
&\Delta F_p(t_k, x(t_k), y(t_k), f(t_k))  = \\ & \;\;\;\;\;\;\;\;\;\;  \begin{cases}  S_{p}\Big( x(t_k), y(t_k), f(t_k) \Big) &  \mbox{if}\,  t_\mathrm{flare} \leq t_k \leq t_\mathrm{flare} + \delta t_\mathrm{flare}
\\ 0 & \mathrm{otherwise,} 
\end{cases}
\end{aligned}
\end{equation}
where $S_p$ is the PRF of the pixel $p$ on the detector and $\delta t_\mathrm{flare}$ is the duration of the flare. For solar-like stars, $\delta t_\mathrm{flare}$ is typical of the order of a few hours \citep{Schrijver2012}. Thus, the model of an image with a flare can be written as follows
\begin{equation}
F_p(t) = S^\star_p(t) +\Delta F_p(t) + n_p(t),  \,\,\, 1\leq p \leq N_\mathrm{pix}, 
\end{equation}
where $n_p(t)$ is the photon noise. 

\begin{figure*}[ht]
\centering
    \includegraphics[width=\textwidth]{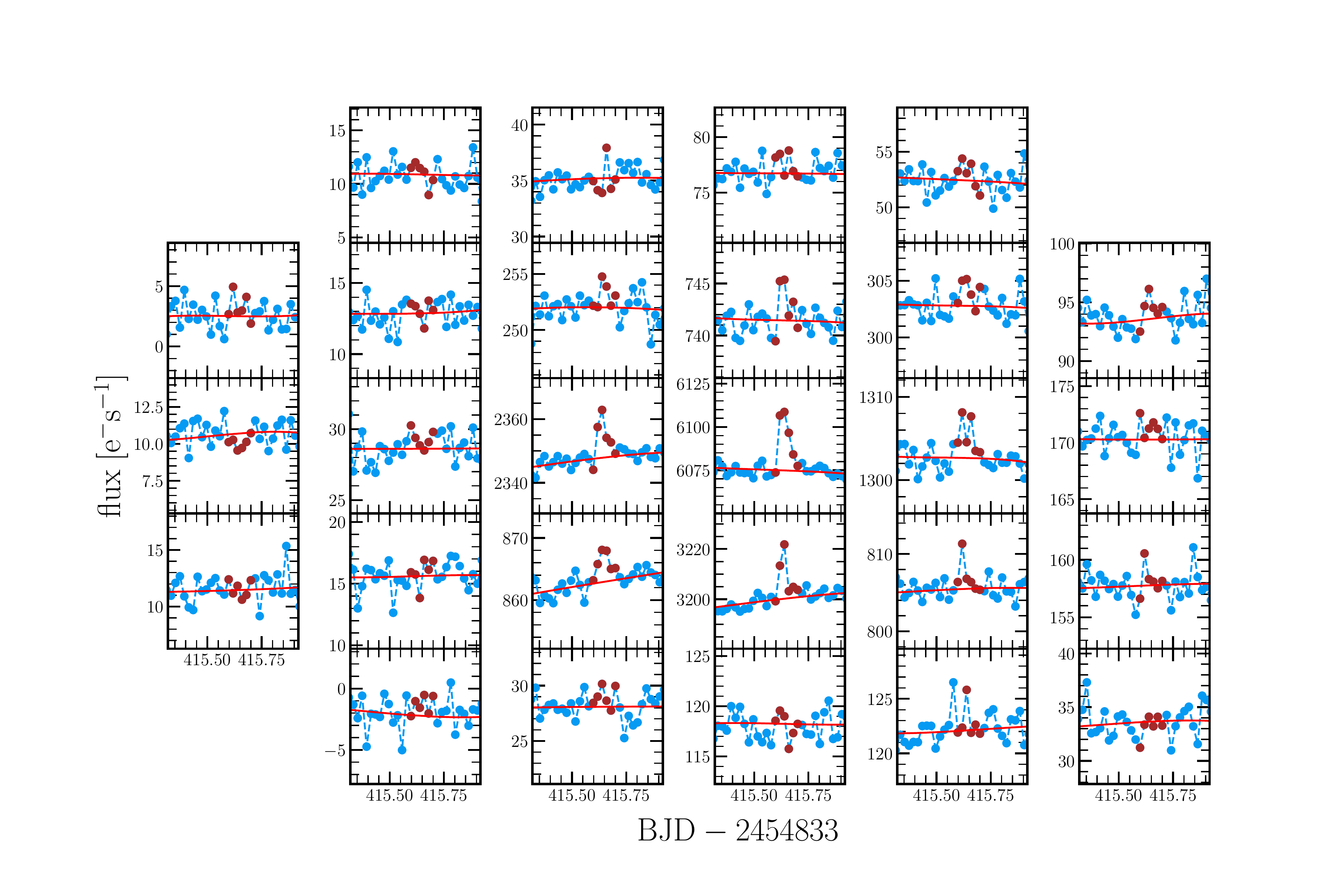}
    \caption{Example of detrending the flux in individual pixels for the star KIC\,10011070 around the flare at $t_\mathrm{flare}=415.62$ days. Blue dots are data points $F_p(t_k)$ used for the model of the flux variation in the absence of the flare $S_p^\star(t_k)$. The data points in all pixels at a given cadence $F(t_k)$ represent an image. Red dots indicate data during the interval $t_\mathrm{flare} \leq t_k \leq t_\mathrm{flare} + \delta t_\mathrm{flare}$, which were excluded from the fitting of $S_p^\star(t_k)$. Note the different scale of the ordinate in individual panels.}
    \label{fig:pixel_light_curve_detrending}
\end{figure*}

The first step is to estimate the quiet stellar flux $S^\star_p(t)$ in each pixel during the flare. For that, we select images in time windows $t \in [t_\mathrm{flare} - \Delta T/2, \,t_\mathrm{flare})$ and $t \in (t_\mathrm{flare}+\delta t_\mathrm{flare}, \,t_\mathrm{flare}+\delta t_\mathrm{flare}+\Delta T/2]$, when the flare contribution is zero, i.e. $\Delta F_p(t)=0$. In these time intervals, we model the stellar flux $S^\star_p(t)$ by a smooth function. Since the brightness does not change significantly on such small timescales, we chose a cubic polynomial to fit the pixel fluxes at the cadences before and after the flare.  In Figure~\ref{fig:pixel_light_curve_detrending}, we show the individual pixel fluxes $F_p(t)$ from a series of images of the target star KIC\,10011070 around $t_\mathrm{flare}=415.62$ days and the fitted model $S^\star_p(t)$ representing the quiet stellar flux.

\begin{figure*}[ht!]
\centering
    \includegraphics[width=\textwidth]{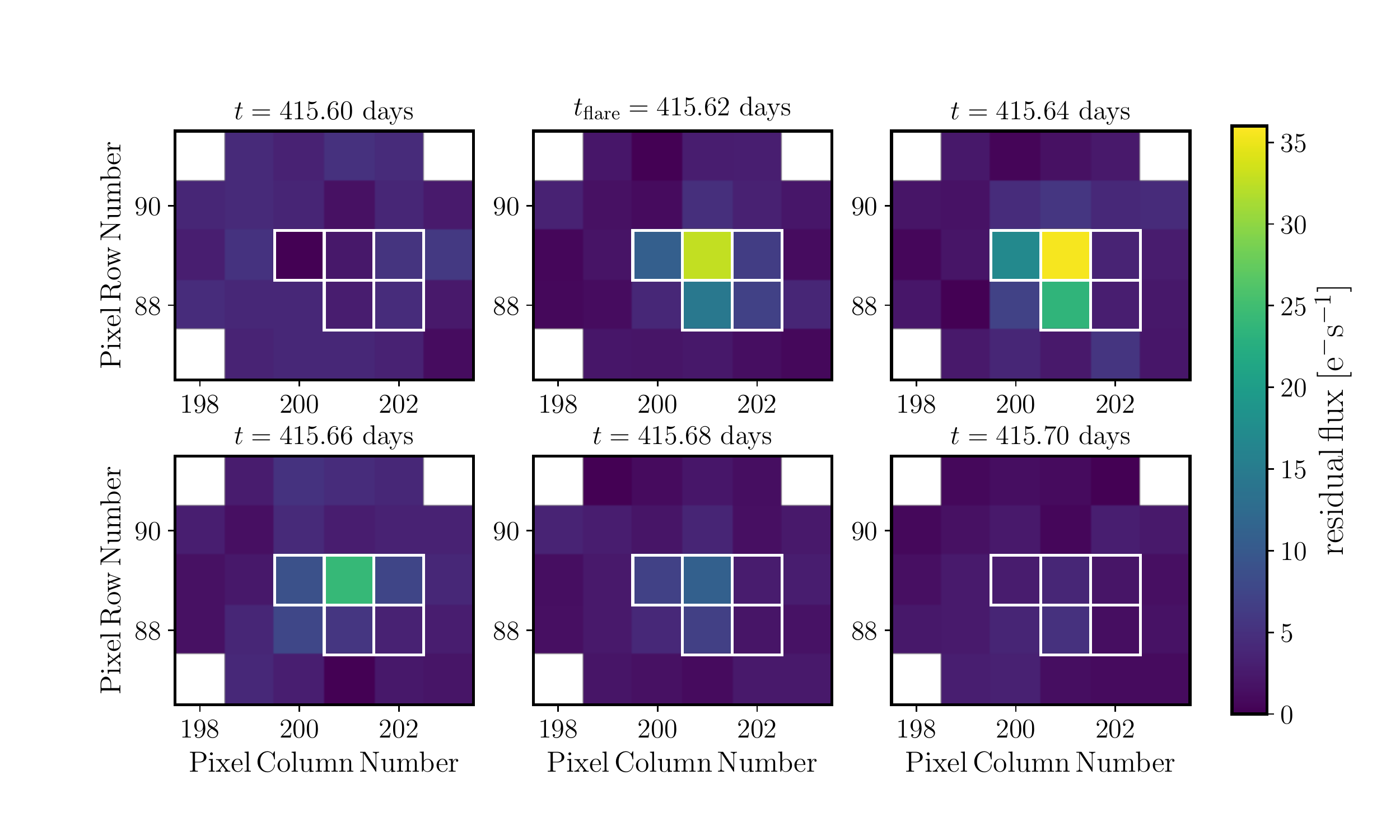}
    \caption{A series of six images $D(t_k)$ with the flare at $t_\mathrm{flare}=415.62$ days for the star KIC\,10011070. The images of the subsequent cadences  $t_\mathrm{flare}=415.62$ and $t_\mathrm{flare}=415.64$ days were used in the PSF-fitting to localize the flare. Squares with white borders show the pixels of the aperture mask used to extract the light curve. }
    \label{fig:residual flares}
\end{figure*}

To obtain a series of images containing only the flux excess from the flare, we subtract the estimated quiet flux from the images. In addition, to avoid negative flux values in the image pixels, we add an artificial offset $c(t_k)$ to each image and take it into account further in the analysis
\begin{equation}
\label{eq:offset}
c(t_k) = \Big| \min_{p} \Big(F_p(t_k)-S^\star_p(t_k) \Big) \Big|.
\end{equation}
The new set of flare images is given by the following expression: 
\begin{equation}
D(t_k) = F(t_k) - S^\star(t_k) + c(t_k),
\end{equation}
where $D(t_k)=\{d_1(t_k), \ldots, d_{N_{\mathrm{pix}}}(t_k)\}$ is a vector representing the new image and in each pixel $p$ the flux $[\mathrm{e}^{-}\mathrm{s}^{-1}]$ is:
\begin{equation}
d_p(t_k)=F_p(t_k) - S^\star_p(t_k) + c(t_k).
\end{equation}
In Figure~\ref{fig:residual flares}, we show a series of six images $D(t_k)$ around $t_\mathrm{flare}$ for the target star KIC\,10011070. 

We introduce a model representing the data: \begin{equation}
d_{p} (t) = \Delta F_p(t, x(t), y(t), f(t)) + b(t) + n_p(t),   \,\,\, 1\leq p \leq N_\mathrm{pix},
\end{equation}
where $b(t)$ is a parameter characterizing the offset that we have introduced in Eq.~\ref{eq:offset}, and  
$\Delta F_p(t, x(t), y(t), f(t))$ is the flux excess due to the flare given by Eq.~\ref{eq:flare_model}. 

For a given cadence $k$, we define a vector of the model parameters as  $\Theta_k= \Big\{x_k, y_k, f_k, b_k \Big\}$.
We estimate them using a Bayesian formalism. The posterior probability distribution $\mathcal{P}(\Theta_k | D_k)$ is the conditional probability of the parameter set $\Theta_k$ given the data $D_k$: 
\begin{equation}
\mathcal{P}(\Theta_k | D_k)=\frac{\mathcal{P}(\Theta_k)}{\mathcal{P}(D_k)} \mathcal{P}(D_k|\Theta_k), 
\end{equation}
where $\mathcal{P}(\Theta_k)$ is the prior probability ascribed to the set of parameters, $\mathcal{P}(D_k|\Theta_k) $ is the likelihood function, and $\mathcal{P}(\Theta_k|D_k)$ is the posterior probability.  $\mathcal{P}(D_k)$ is a normalizing factor that can be ignored for our purposes. 

We assume that photons and electron counts in pixels obey Poisson statistics. Therefore, we use the  likelihood function in the form of the Poisson distribution: 
\begin{equation}
\mathcal{P}(D_k|\Theta_k) = \prod_{p=1}^{N_\mathrm{pix}} \exp\Big(-\lambda_{p}(\Theta_k)\Big) \frac{1}{d_{p}!}  \lambda_{p}(\Theta_k)^{d_{p}}, 
\end{equation}
where $\lambda_p(\Theta_k)$ is the expected flux for the $p$-th pixel given by the model:
\begin{equation}
\lambda_p(\Theta_k) = \Delta F_p(x_k, y_k, f_k) + b_k. 
\end{equation}

Taking into account the independence of parameters, one can write the prior probability as the product
\begin{equation}
\mathcal{P}(\Theta_k) = \prod_{\alpha=1}^{N_{\Theta}} \mathcal{P}(\Theta_{k, \alpha}), 
\end{equation}
where $N_\Theta=4$ is the number of the fitted parameters. 
For the flare flux $f_k$,  we take the uniformly distributed prior
\begin{equation}
\mathcal{P}(f_k) = \mathcal{U}(0, f_{k,\mathrm{ max}}), 
\end{equation}
where $f_{k,\mathrm{max}}=\sum_{p=1} ^{N_\mathrm{pix}} d_p(t_k)$ and $\mathcal{U}$ denotes the uniform distribution. For the flare location, we take the uniformly distributed priors:
\begin{equation}
\mathcal{P}(x_k) = \mathcal{U}(x_{k,\mathrm{ min}}, x_{k,\mathrm{ min}} + N_\mathrm{col} )
\end{equation}
and 
\begin{equation}
\mathcal{P}(y_k) = \mathcal{U}(y_{k,\mathrm{ min}}, y_{k,\mathrm{ min}} + N_\mathrm{row} ),
\end{equation}
where $x_{k,\mathrm{ min}}$ and  $y_{k,\mathrm{ min}}$ are the pixel coordinates of the bottom left corner of the image,  $N_\mathrm{col}$ and $N_\mathrm{row}$ are the number of columns and rows of the image.

For the offset, we take the Gaussian prior 
\begin{equation}
\mathcal{P}(b_k) = \frac{1}{ \sqrt{2\pi \tilde{\sigma}^2_{b} }} \exp \Big(  -\frac{1}{2} (b_k - \mu_{k,b})^2/\tilde{\sigma}^2_{k,b} \Big)
\end{equation}
with the width $\tilde{\sigma}_{k,b}^2 = {\rm Var}[D(t_k)] $ and the mean 
\begin{equation}
\mu_{k,b} = \sum_{p=1} ^{N_\mathrm{pix}} d_p(t_k)/N_\mathrm{pix}.
\end{equation} 

%%%%%%%%%%%%%%%%%%%%%%%%%%%%%%%%%%%%%%%%%%%%%%%%
Since we analyze Kepler data, we use the PRF of the Kepler telescope but we note that the proposed method can be adapted for any photometric survey with known PRF. We use the \textsc{Lightkurve} package  \citep{2018ascl.soft12013L} to compute theoretical pixel fluxes $\lambda_{p} \Big(x_k, y_k, f_k\Big)$ on an image assuming a point source with a given position $(x_k,y_k)$ and flux $f_k$. 

We separately analyze two images $D(t_\mathrm{flare})$ and $D(t_\mathrm{flare} + \Delta t)$. According to the given threshold, there is a flux excess above $5\sigma$ in the light curve for these two cadences. We use the Goodman \& Weare’s Affine Invariant Markov chain Monte Carlo (MCMC) Ensemble sampler \textsc{emcee} \citep{Goodman2010, 2013PASP..125..306F} to compute the marginalized probability functions. A vector with initial parameters we denote as  $\Theta_k^{(o)}= \Big\{x_k^{(o)}, y_k^{(o)}, f_k^{(o)}, b_k^{(o)} \Big\} $. The center of brightness of the image $D(t_k)$ we use as the initial flare position:
\begin{equation}
x_k^{(o)} = \sum_{p=1}^{N_\mathrm{pix}} X_p d_p(t_k)/ \sum_{p=1}^{N_\mathrm{pix}}  d_p(t_k)
\end{equation}
and 
\begin{equation}
y_k^{(o)} = \sum_{p=1}^{N_\mathrm{pix}} Y_p d_p(t_k)/ \sum_{p=1}^{N_\mathrm{pix}}  d_p(t_k),
\end{equation}
where $(X_p, Y_p)$ is the $p$-th pixel center coordinates.
The sum of the pixel fluxes we take as an initial guess of the flare flux:
\begin{equation}
f_k^{(o)} = \sum_{p=1}^{N_\mathrm{pix}} d_p(t_k).
\end{equation}
The mean of the pixel fluxes is taken as an initial guess of the offset:
\begin{equation}
b_k^{(o)} = \sum_{p=1}^{N_\mathrm{pix}} d_p(t_k)/N_\mathrm{pix}.
\end{equation}
From the MCMC fitting, we get the final marginalized probability distribution functions (PDFs) of the model parameters.

The implementation of the method performing the steps described above is available in \textsc{python}.  We provide an open source package \textit{Localization Of Superflare Events (\textsc{LOSE})} on GitHub\footnote{\url{https://github.com/ValeriyVasilyevAstro/LOSE}}.

\subsection{Flare energies}
For flares occurring on the target star, we measure the flare energy $E_\mathrm{flare}$. We assume that the flare radiates as a black body with an effective temperature $T_\mathrm{flare} = 10,000$\,K  \citep{Hawley1992, Shibayama2013, Guenther2020}. 
The energy emitted by the flare each second is its bolometric luminosity:
\begin{equation}
L_\mathrm{flare}(t, T_\mathrm{flare}) =  A_\mathrm{flare}(t) \int B_\lambda (T_\mathrm{flare}) d\lambda,
\label{eq:lum_flare}
\end{equation}
where $A_\mathrm{flare}(t)$ is the area of the flare and $B_\lambda (T_\mathrm{flare})$ the black body spectrum integrated over all wavelengths.
The total energy emitted by a flare is the integral of the luminosity over the flare duration
\begin{equation}
\label{eq:flareenergy}
E_\mathrm{flare} = \int_{t_\mathrm{flare}}^{t_\mathrm{flare} +\delta t_\mathrm{flare} } L_\mathrm{flare}(t, T_\mathrm{flare}) dt. 
\end{equation}

The signal in each pixel of the CCD-detector with a known transmission function $\phi(\lambda)$ is proportional to the number of photons falling onto the pixel during the exposure time $\Delta t$. Therefore, to obtain an estimate of the bolometric flare energy one has to convert the photon luminosity of the flare, denoted as $N_\mathrm{flare}$, into the flare  bolometric luminosity. The number of photons emitted by the flare at each time $t$ in the passband $\phi(\lambda)$ is 
\begin{equation}
N_\mathrm{flare}(t, T_\mathrm{flare}) =  A_\mathrm{flare}(t) \int \frac{ \lambda B_\lambda (T_\mathrm{flare})}{hc} \phi(\lambda) d\lambda,
\label{eq:phot_lum_flare}
\end{equation}
where $c$ is the speed of light, and $h$ is the Planck constant.
Then, the conversion factor $\alpha(T_\mathrm{flare})$  is given by the time-independent ratio
\begin{equation}
\alpha(T_\mathrm{flare}) = L_\mathrm{flare} (t, T_\mathrm{flare}) / N_\mathrm{flare} (t, T_\mathrm{flare}).
\label{eq:alphaflare}
\end{equation}
Using Eqs.~\ref{eq:lum_flare} and \ref{eq:phot_lum_flare}, we rewrite the ratio as:
\begin{equation}
\alpha(T_\mathrm{flare}) = \frac{\int B_\lambda (T_\mathrm{flare}) d\lambda}{  \int \frac{ \lambda B_\lambda (T_\mathrm{flare})}{hc} \phi(\lambda) d\lambda}.
\label{eq:alphaflare2}
\end{equation}
In a similar way, we define a conversion factor $\alpha(T_*)$ between the total number of emitted photons and the total emitted energy for the star as a whole:
\begin{equation}
\alpha(T_*) = \frac{\int B_\lambda (T_*) d\lambda}{  \int \frac{ \lambda B_\lambda (T_*)}{hc} \phi(\lambda) d\lambda},
\end{equation}
where $T_*$ is the effective temperature, and $N_*$ is the number of emitted photons of the star.

In the light curve, the flare appears as a flux excess
\begin{equation}
\Delta l_\mathrm{flare}(t) \sim N_\mathrm{flare}(t, T_\mathrm{flare}) /r_*^2
\end{equation} 
over the quiet stellar flux:
\begin{equation}
\label{eq:lcstar}
l_*  \sim N_*(T_*)/r_*^2, 
\end{equation}
where $r_*$ is the distance from the observer to the star.  Then,  by combining Eqs.~\ref{eq:alphaflare2}--\ref{eq:lcstar}, we obtain the relationship between the bolometric luminosity of the flare and the observed light curve:
\begin{equation}
\frac{L_\mathrm{flare}(t, T_\mathrm{flare})}{L_*(T_*)} = \frac{ \alpha(T_\mathrm{flare})  }{ \alpha(T_*) }  \frac{\Delta l (t)}{l_*},
\end{equation} 
which we use with Eq.~\ref{eq:flareenergy} to estimate the flare energy. 

\section{Results}
\subsection{Light curve analysis}
From the flare search in the light curves (see Sect.~\ref{sec:lightcurves}), we obtained a list of flare candidates. In total, we found \NLC cases where at least two consecutive data points lie above the $5\sigma$ level in the detrended light curves. It is important to note that this number most likely also contains events not associated with the target star (e.g. cosmic rays, minor planets, residual instrumental effects etc.). For each event that has been detected in the light curve, we performed the PSF-fitting (see Sect.~\ref{sec: flare_localization}).

\subsection{PSF-fitting of the flare} \label{sec:psf-results}
\begin{figure}[ht!]
\centering
\includegraphics[width=0.5\textwidth]{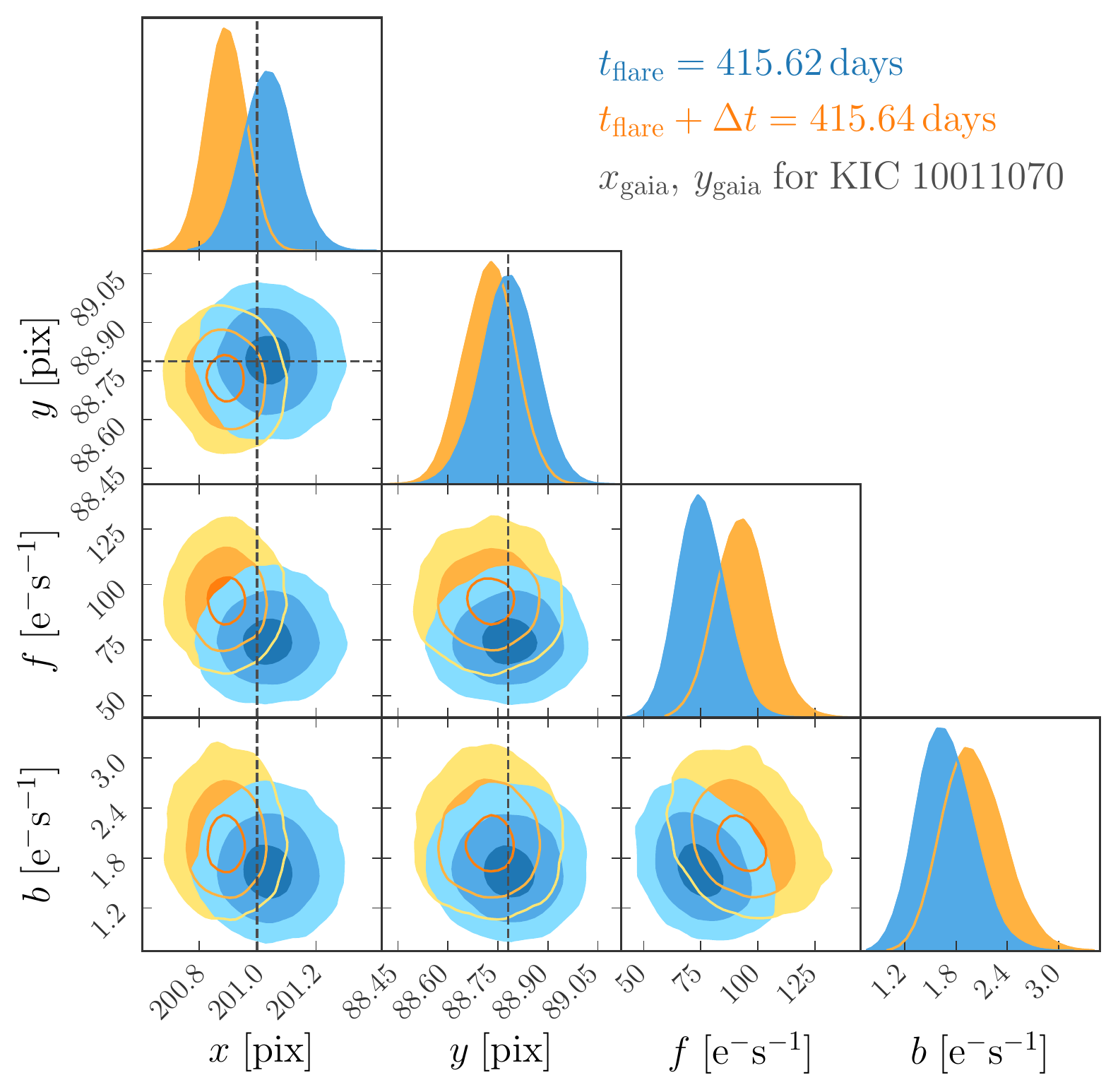}
\caption{Marginalized 1D and 2D PDFs of the model parameters used to fit the flare to the images $D(t_\mathrm{flare})$ and $D(t_\mathrm{flare} + \Delta t)$. Blue and orange colors denote the distributions derived by separately fitting the two cadences. In the 2D plots, the different shades of blue and orange represent the 68\%, 95\%, and 99.9\% confidence contours. The black dashed lines represent the pixel coordinates of the target star KIC\,10011070 obtained by the coordinate transformation of the Gaia DR2 coordinates $(\alpha, \delta)$. }
\label{fig:mcmc_marginalized_distributions}
\end{figure}

With the MCMC fitting, we obtained the  one-dimensional marginalized PDFs of the fit parameters.  In Figure~\ref{fig:mcmc_marginalized_distributions}, we show the marginalized 1D and 2D PDFs for the flare position $(x(t_k), y(t_k))$, the flare flux $f(t_k)$, and the offset $b(t_k)$ for the two subsequent cadences $t_k = t_\mathrm{flare}$ and $t_k = t_\mathrm{flare} + \Delta t$ of the star KIC\,10011070. 

The position of the flare is constrained by drawing 2D  confidence regions on the image. For that, we take into account that the sum of squares of two normally-distributed random variables, $x$ and $y$, has a chi-square distribution with two degrees of freedom. Based on that, we draw  68\%, 95\%, and 99.9\% confidence ellipses.

The PSF-fitting returns the distributions of the most likely positions of the flare candidate for the two cadences. To match the location of the target (or any other star) and the confidence regions, we use the Gaia DR2 catalog \citep{GaiaDR2_2018}. We extract the sky coordinates $(\alpha, \delta)$ of all stars (including the target star) within a radius of 10 arcsec around the target star, and convert them into the pixel coordinates using the World Coordinate System (WCS) transformation. For both cadences, we check whether a background or the target star lies within the 99.9\% confidence level ellipse. To associate the potential flare with the star (target or background) we require that the star position is within the 99.9\% confidence ellipse for both cadences.
\begin{figure}[ht!]
\centering
\includegraphics[width=0.45\textwidth]{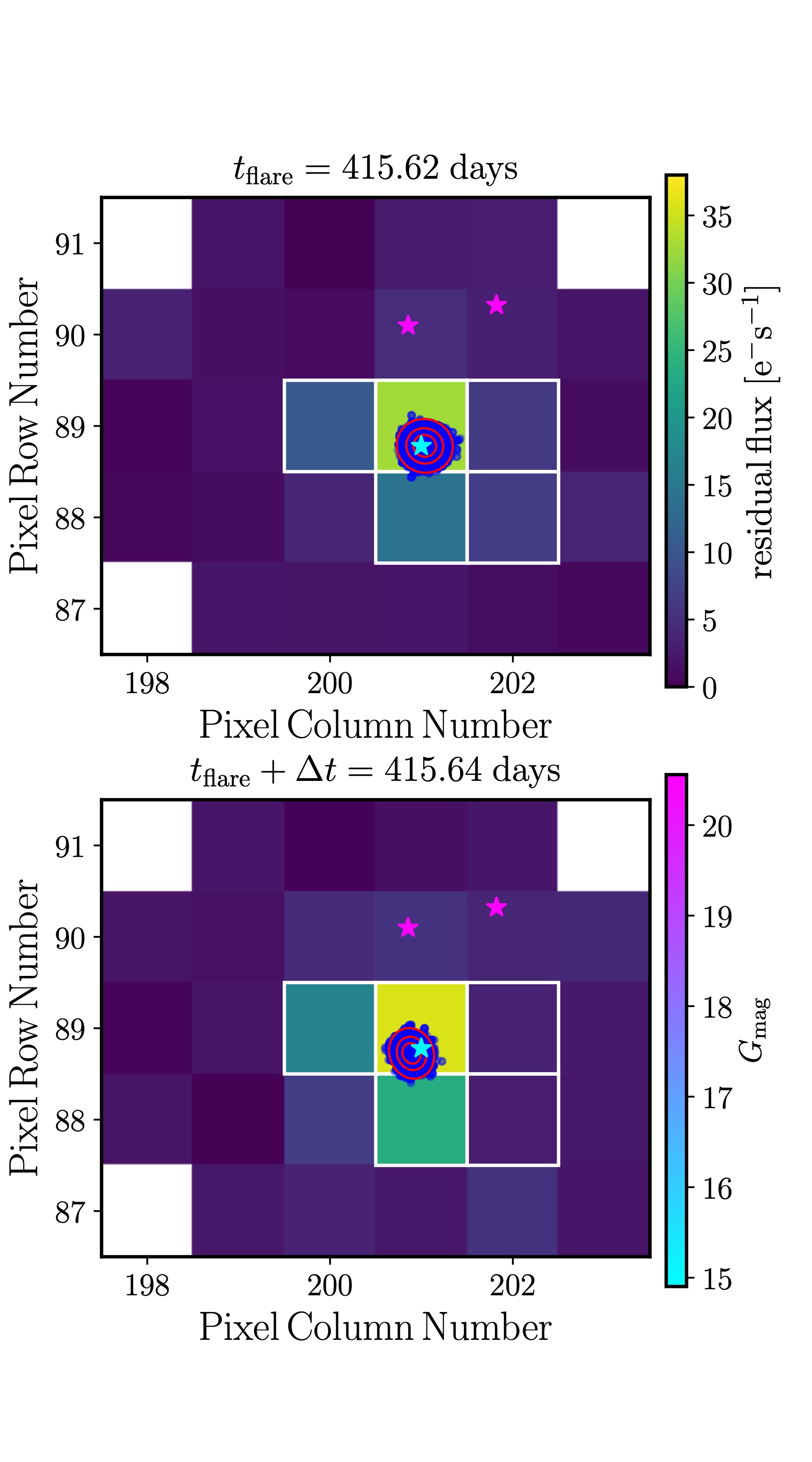}
    \caption{Example of the flare localization in the target pixel files for the two subsequent cadences $t_\mathrm{flare}$ (top) and $t_\mathrm{flare} + \Delta t$ (bottom). The star symbols are color-coded according to their apparent Gaia magnitude $G_\mathrm{mag}$. 
    % The target star KIC\,10011070 is in cyan and two background stars are in pink. 
    The blue dots represent 1000 realizations of the flare position obtained from the MCMC fitting, and the red ellipses show the 68\%, 95\%, and 99.9\% confidence regions.  Squares with white borders show the pixels of the aperture mask used to extract the light curve. }
    \label{fig:ellipces_on_the_sky_with_the_gaia_stars}
\end{figure}

In Figure~\ref{fig:ellipces_on_the_sky_with_the_gaia_stars}, we show the 2D confidence regions of the flare position for the target star KIC\,10011070 as well as stars from the Gaia DR2 catalog on the two images $D(t_\mathrm{flare})$ and $D(t_\mathrm{flare} + \Delta t)$. In the regions of interest, we found three stars in the Gaia DR2 catalog: the target star and two fainter background stars. For both cadences, only the target star is located inside the 99.9\% confidence region, whereas both contaminant stars are far away from the estimated flare location. This leads us to conclude that the flare most likely happened on the target star.

It is important to note that we found cases when the confidence ellipse with the possible flare position is much bigger than the pixel size. Such cases indicate that a proper solution in the MCMC runs cannot be found and that the observed signal is not associated with a flare. This is, for example, the case for the transits of minor planets. Therefore, we retain only  flare candidates when the possible flare position derived from the MCMC fit satisfies the following condition (for both cadences)
\begin{equation} 
 a(t_k)+g(t_k) < 2 \, \mathrm{pixels}, 
\label{ellipce_condition_}
\end{equation}
where $a(t_k)$ and $g(t_k)$ are the semi-axes of the 99.9\% confidence level ellipse for $t_k=t_\mathrm{flare}$ and $t_k=t_\mathrm{flare} + \Delta t$. The threshold of 2 pixels is taken to reduce the number of cases when the ellipse size becomes comparable with the typical size of the photometric mask used to extract the light curve.

These two criteria, the location of a star within the 99.9\% confidence region for both cadences and the sizes of the confidence ellipses given by Eq.~\ref{ellipce_condition_}, reduce the number of potential flares to 374. Next, for these events, we  visually inspected the flare profiles in the light curves. Typically, the flare profile shows a steep increase of the flux followed by an exponential decay. However, we have found several cases when the flare profile has a symmetric shape with four or more points above the $1\sigma$ level. We associate these events with minor solar system objects (see Sect.~\ref{sec:not_flares}). In addition, we found several cases when the flare profile in the light curve has a long rising phase ($\ge1.5$ hours above the $1\sigma$ level) and a short descending phase ($\sim30$ min). We discarded such events from further analysis. The visual inspection further reduced the number of potential flares to 361. 

\subsection{Distribution of events in time} \label{sec:time-ditribition-results}
\begin{figure}[h]
\centering
\includegraphics[width=0.49\textwidth]{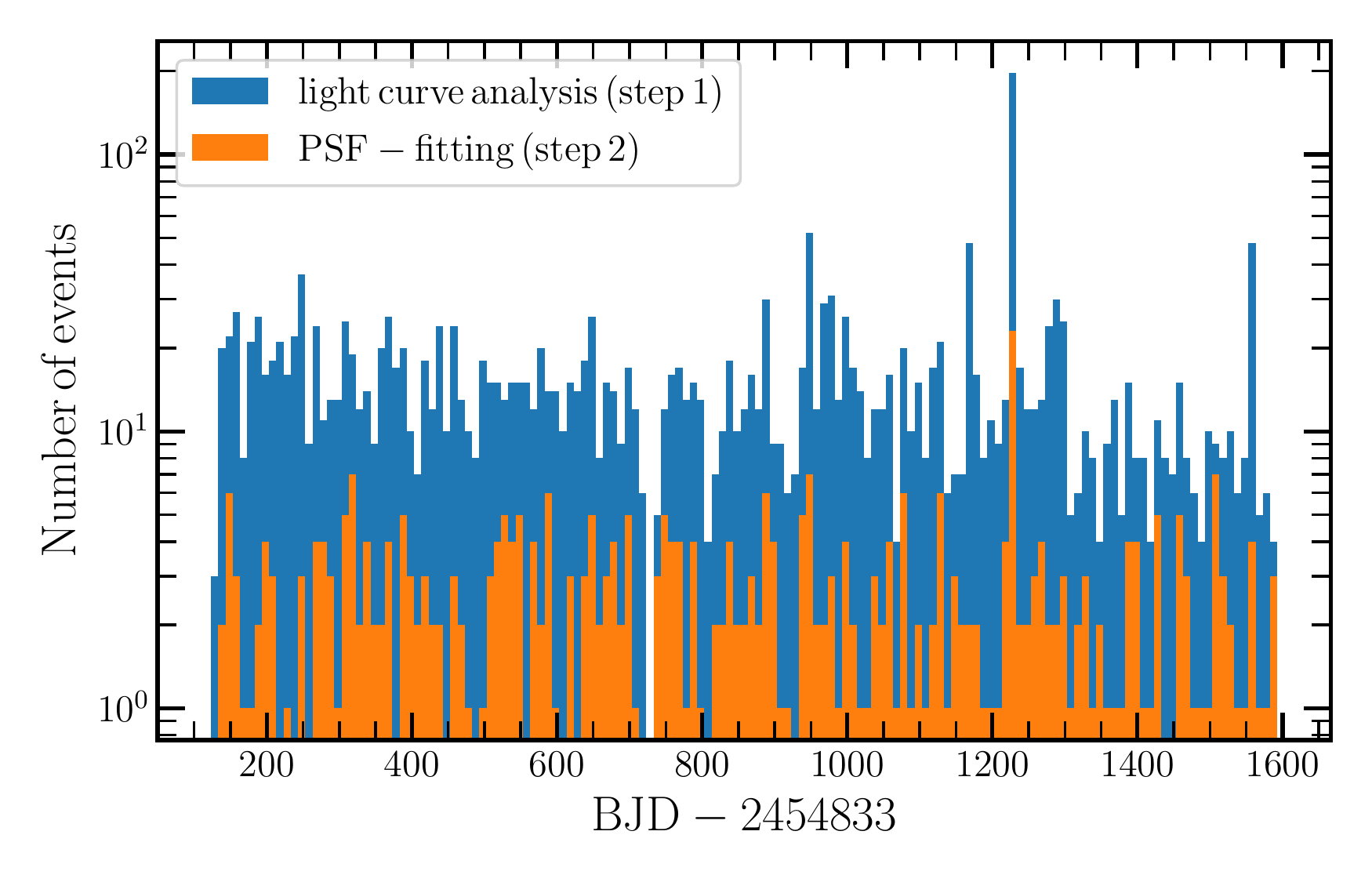}
\caption{Distribution of possible flare events in time after the analysis of the light curves (blue) and the PSF-fitting of the flare images (orange). The bin width is 10 days.}
    \label{fig:distribution_in_time}
\end{figure}

A flare observed on a certain star happens independently from flares occurring on other stars. Thus, one would expect a uniform distribution of events in time. However, the presence of residual instrumental effects (i.e., an increase of the instrumental noise level in a short time interval) can produce outliers erroneously interpreted as flares. Therefore, we checked the temporal distribution of events detected in the light curve analysis (\NLC events) and those after the flare localization using the PSF-fitting method (361 events). These distributions are shown in Figure~\ref{fig:distribution_in_time}. The distribution of events from the light curve analysis appears non-uniform. Around the times $t=1231$ days and $t=1561$ days, there have been detected an anomalously large number of flare candidates. After the PSF-fitting procedure and further analysis, the peak at $t=1561$ days vanished but the peak at $t=1231$ days persists in the distribution. The inspection of the individual pixel fluxes showed that there was a systematic increase in the noise at that time. This increased noise level can be seen in the light curve and in the target pixel files. During the period $t=1231-1233$ days we found 19 events, which have been excluded from further analysis. Thus, our final list of flares associated with the target (or the background) star consists of \NF~events.

\subsection{Classification of events}\label{sec:classification_of_events}
Based on the results of the PSF-fitting, we classify all events into the following groups: (i) flare on the target star,  (ii) flare on a background star, (iii) flare that cannot be distinguished between the target and a background star, and (iv) (likely) not flares. 
In Table~\ref{tab:final_summary_with_the_detected_events}, we show the summary with the number of events detected in each step, and the number of flares in each group. The groups are described in detail below.

\begin{table*}[ht!]
\centering
\begin{tabular}{llr}
\hline
    & Number of events & Number of stars  \\
\hline
 Solar-like stars in the input sample & - & 5862  \\
Events detected in the light curves &  \NLC & 1696 \\
Flares detected with the PSF-fitting & \NF & 208\\
Flares detected on the target star & \NFTS & \NSTS \\
Flares detected on a background star & \NFBS & \NSBS\\
Flares indistinguishable between  the target and a background star & \NFBOTS & \NSBOTS \\
\hline
\end{tabular}
\caption{Summary of detected events in the light curves and confirmed or rejected by the PSF-fitting. The number of stars associated with these events are given in the right column.}
\label{tab:final_summary_with_the_detected_events}

\end{table*}

In Table~\ref{tab:final_list_of_events}, we list all detected flares and provide the Kepler ID of the target star (KIC), the Gaia DR2 ID, the time $t_\mathrm{flare}$ of the first cadence above the $5\sigma$ threshold,  the flare duration $\delta t_\mathrm{flare}$, and the stellar rotational period from \citet{McQuillan2014} if available. Additionally, we introduce a flag to distinguish between the three groups of flares. We set flag=1 for flares detected on the target star, flag=2 for flares detected on a background star, and flag=3 if the flare cannot be distinguished between the target and background star. For the first group, we also provide the flare energy $E_{\rm flare}$.  For the last two cases, we provide the Gaia DR2 ID for the background star. A machine-readable version of the table is available at the CDS.
% The full table with all detected flares is available in the online-only format. 

\begin{table*}[ht!]
\centering
\begin{tabular}{lllllllll}
\hline
\hline
KIC ID & Gaia DR2 ID  & $t_\mathrm{flare}$  & $\delta t_\mathrm{flare}$ & $E_\mathrm{flare}$ & $P_\mathrm{rot}$ &  flag &  Background star Gaia DR2 ID \\ 
& & [BJD - 2454833] & [min] & [erg] & [days] & & \\
\hline
3120315 & 2051936538528690560 & 950.024 & 49.98  & $3.73 \cdot 10^{34}$ & - & 1 & \\
3235322 & 2052679980185318528 & 941.809  & 48.70 & $1.02 \cdot 10^{35}$ & - & 1 &  \\
\dots \\
\hline

\end{tabular}
\caption{The catalog of detected flares. Full is available in electronic form at the CDS. }
\label{tab:final_list_of_events}
\end{table*}

\subsubsection{Flares on the target star}
This group contains all events for which the target star lies within the 99.9\% confidence ellipse for both cadences. In Figure~\ref{fig:flare_on_the_target_star}, we show several examples with flares in the light curve and the images. Here, we also include cases when both the target and a background star are contained within the ellipse at one cadence, but only the target star lies in the ellipse at the other cadence (see the top panel in Fig.~\ref{fig:flare_on_the_target_star}). In total, this group includes \NFTS flares on \NSTS solar-like stars.  Among the \NSTS stars, for 22 stars rotation periods have been measured by \citet{McQuillan2014}. For all \NFTS flares, we estimate the flare energy and its duration. In Table~\ref{tab:final_list_of_events}, we set flag=1 for these events. Figures with light curves and results of the flare localization for these events can be found on the GitHub page of the project \footnote{\url{https://github.com/ValeriyVasilyevAstro/LOSE/results/}}.

\subsubsection{Flares on background stars}
This group consists of the events for which at least one background star lies within the 99.9\% confidence ellipse and the target star is not located in the confidence ellipse for both cadences $t_\mathrm{flare}$ and $t_\mathrm{flare}+\Delta t$. We show examples in  Figure~\ref{fig:flare_on_background_star}. 
In this group, the largest number of flares (6 in total) is detected in the light curve data of KIC\,7189661. However, according to the PSF-fitting, these flares occurred on a faint background star ($m_G=20.3$) with the catalog number Gaia DR2\,2102905052858466816. In total, we found  \NFBS flares on  \NSBS background stars in this group. In Table~\ref{tab:final_list_of_events}, we set flag=2.

\subsubsection{Flares indistinguishable between the target and a background star}
This group refers to cases when the target star and the same background star are within the confidence ellipse for both cadences, $t_\mathrm{flare}$ and $t_\mathrm{flare}+\Delta t$. Thus, there is no possibility of uniquely assigning the flare to the target or the background star. We show several examples in Figure~\ref{fig:multiple_stars_in_ellipce}. In this group, we found \NFBOTS events on  \NSBOTS stars. In Table~\ref{tab:final_list_of_events}, we set flag=3.

\subsection{Rejected cases}
\label{sec:not_flares}
This group encompasses events that have at least two points above the $5\sigma$ threshold in the light curve and belong to one of the following cases: 
(i) they do not satisfy the condition constraining the size of the confidence ellipse given by Eq.~\ref{ellipce_condition_} for both cadences,  (ii) the star (either target or background) within the confidence ellipse is not the same for both cadences, (iii) the flare profiles with more than 3 points above the $5\sigma$ threshold in the light curve are symmetric or have longer rising phase compared to the descending phase. In Figure~\ref{fig:ellipce_different_location}, we present several examples of such events.

Different physical processes can cause these types of events. Potentially, it can be a transit of a solar-system object across the image (see top panel  Fig.\ref{fig:ellipce_different_location}). Typically, such transits show a symmetric shape in the light curve, i.e., the rising and declining phases look quite similar. Owing to the fast proper motion of such objects relative to the stars, their reflected light is generally distributed over more pixels than the light from a single star (even during a single exposure). This effectively increases the size of the 2D confidence region, i.e., the ellipse size constraining the location of the event. 

In addition, we found events that consist of two points above the $5\sigma$ threshold in the light curve but the confidence ellipses for the two subsequent cadences are located in two different pixels and the center of one ellipse lies outside of the other one (see the middle panel in Fig.\ref{fig:ellipce_different_location}). Possible sources of such events are  subsequent cosmic ray hits. Cases were also found for which the location of the confidence ellipse does not change from one point in time to the other, but it does not overlap with any star (see the bottom panel in Fig.\ref{fig:ellipce_different_location}).
Such events might be caused by other astrophysical sources too faint to be detected by Gaia.

\subsection{Comparison to Okamoto et al. (2021)}

In this section, we compare our flare detections to those reported in the literature. Here, we focus on the latest flare catalog of solar-like stars provided by \citet{Okamoto2021} (hereafter \citetalias{Okamoto2021}).

\citetalias{Okamoto2021}  conducted a statistical analysis of superflares on solar-like main-sequence stars. From the Kepler light curves, these authors detected 2344 flares on 256 stars with rotational periods longer than one day up to 40 days. Among these stars, there is a sub-sample of 16 solar-like stars with $T_\mathrm{eff}=5600-6000$\,K and $P_\mathrm{rot}>20$ days (see Table~2 in \citetalias{Okamoto2021}). To these stars, we applied our two-step algorithm described in Section~\ref{sec:method} and found 38 flares. Interestingly, \citetalias{Okamoto2021} identified 29 flares. From our 38 flares, 22 flares coincide with those from O2021, and we identified 16 flares not listed by \citetalias{Okamoto2021}.
In addition, we applied our PSF-fitting method to localize the 29 flares reported by \citetalias{Okamoto2021} using our PSF-fitting method. As a result,  7  events have been rejected by the PSF-fitting because the target star was not within the 99.9\% confidence ellipse. In Fig.~\ref{fig:okamoto_stars}, we show the light curves and images for these 7 rejected cases.

\subsection{Flare energies}
\begin{figure*}[ht!]
\centering
\includegraphics[width=0.95\textwidth]{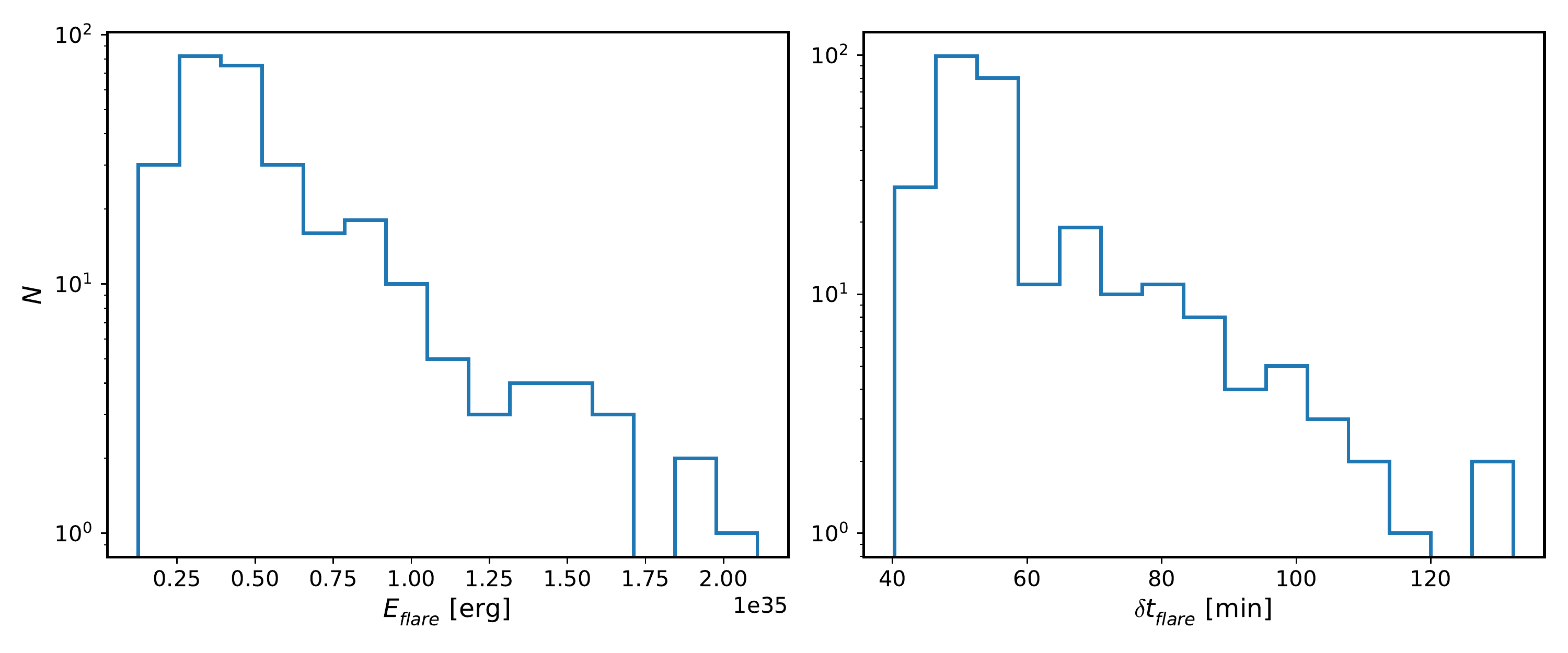}
\caption{Distribution of energies $E_\mathrm{flare}$ (left) and duration $\delta t_\mathrm{flare}$ (right) of flares associated with  the target star.}
    \label{fig:flare_energies_and_duration}
\end{figure*}

Figure~\ref{fig:flare_energies_and_duration}  shows the distribution of flare energies and durations for all events which we found to be associated with a target star. The flare energies are comparable with the estimates presented in the literature \citep{Shibayama2013, Maehara2012, Okamoto2021}. The highest flare energy measured in our sample is around $E_\mathrm{flare}\approx 2 \times 10^{35}$, which is roughly twice the maximum flare energy reported by \citetalias{Okamoto2021}. The difference can potentiall be due to the photon number to energy conversion factors, which has not been used in the literature. For stars from our sample with effective temperatures $T_*=5500-6000$\,K, we estimate  $\alpha(T_\mathrm{flare})/\alpha (T_*) \sim 1.5$.

The proposed method for searching flares in the light curves, that is, the condition of having at least two subsequent data points above the $5\sigma$ threshold, sets the lower limit on the flare duration at $\delta t_\mathrm{min} > 30$ min. Together with the assumption on the flare temperature, it also sets the lower limit of the flare energy, to which the method is sensitive. 

\section{Conclusions}
In this work, we presented a new method to identify the flare location in the target pixel files using the PSF-fitting procedure. Previous flare detection methods were mostly based on the analysis of the light curves only. Our light curve analysis detected \NLC cases that could potentially be interpreted as a flare on the target star. However, the PSF-fitting of the pixel data revealed that for more than 1900 of these cases the target star was not enclosed in the ellipses for both cadences $t_\mathrm{flare}$ and  $t_\mathrm{flare} + \Delta t$. Thus, the flux excess detected in the light curve might arise from some other source of contamination, and cannot be uniquely attributed to the target star.

Even after the PSF-fitting of the pixel data, not all detected flares can be associated with the target star. Of the \NF detected flares, only \NFTS actually happened on the \NSTS target stars, while the remaining cases either happened on a fainter background star or cannot be uniquely attributed. Out of these \NFTS events, 45 flares happened on 22 stars for which rotational periods have been measured by \cite{McQuillan2014}, and 238 flares occurred on 156 solar-like stars with unknown rotation periods. However, we cannot completely exclude the possibility that some of these 156 stars exhibit fast-to-medium rotation period stars (e.g., $P_\mathrm{rot}<20$ days), which have been missed in the automated survey of \citet{McQuillan2014}. This would, in turn, explain some of the flare detections. 
To test this hypothesis, further spectroscopic observations of all 178 flare stars are needed because previous spectroscopic characterizations of Sun-like superflare stars (e.g., \citealt{Notsu2019}) are scarce (see Appendix A in \citealt{Okamoto2021}).

To avoid false detections of flares on neighboring background stars, previous studies selected isolated stars (e.g., \citealp{Okamoto2021}). This criterion, however, often removes more than half of the stars from the initial target list (e.g., \citealt{Shibayama2013} removed 70\% from the initial sample). Our method does not require any sample reduction and allows us to detect flares on stars originally rejected in previous studies. That might help to improve statistics and provide a better estimate of the superflare occurrence frequency on Sun-like stars.

The results of our analysis lead us to conclude that the analysis of light curves alone is insufficient to associate outliers in the time series with flares on the target star. Thus, the flare statistics of previous studies might be contaminated by instrumental effects or unresolved background stars.

We emphasize that the method can, in principle, be applied to any photometric survey with known PSF. Thus, we expect the method to be useful for the analysis of the data from the TESS \citep{TESS} and upcoming PLATO missions \citep{PLATO2}.

\begin{acknowledgements}
VV acknowledges support from the Max Planck Society under the grant ``PLATO Science'' and from the German Aerospace Center under  ``PLATO Data Center'' grant   50OO1501.
      Part of this work was supported by the German
      \emph{Deut\-sche For\-schungs\-ge\-mein\-schaft, DFG\/} project
      number Ts~17/2--1. 
      Part of this work has received funding from the European Research Council (ERC) under the European Union's Horizon 2020 research and innovation program (grant agreement No. 715947).
      IU acknowledges support by the Academy of Finland (grant no. 321882 ESPERA), and inspiring discussions in the framework of the ISSI International team \#510 SEESUP.
LG acknowledges support from ERC Synergy Grant WHOLE SUN 810218. We also  gratefully acknowledge use of the open-source  codes  \textsc{matplotlib} \citep{2007CSE.....9...90H}, \textsc{numpy} \citep{5725236}, \textsc{scipy} \citep{scipy}, and \textsc{pygtc} \citep{Bocquet2016}. 
\end{acknowledgements}

\bibliographystyle{aa} 
\bibliography{bbl}

\begin{appendix}
% In this section, we present different events that have been rejected after the result of the PSF-fitting.
% \newpage
\section{Flare on the target star.}
\begin{figure*}[ht!]
        \centering
              \includegraphics[width=\textwidth]{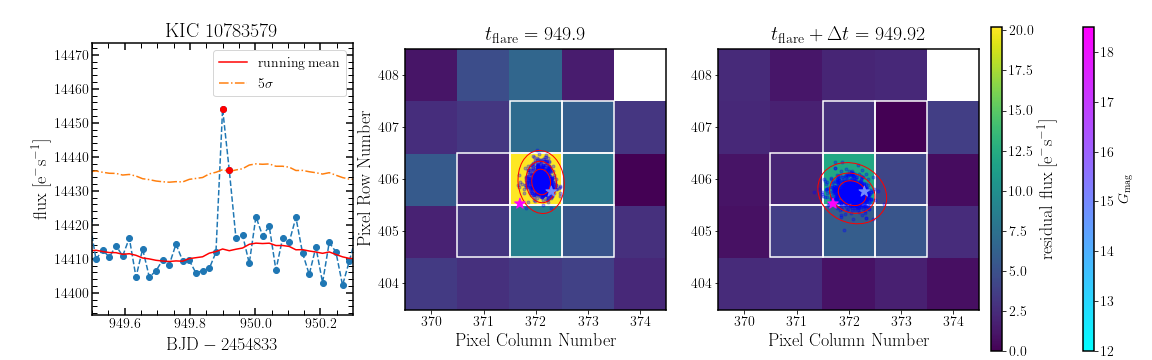}
                \includegraphics[width=\textwidth]{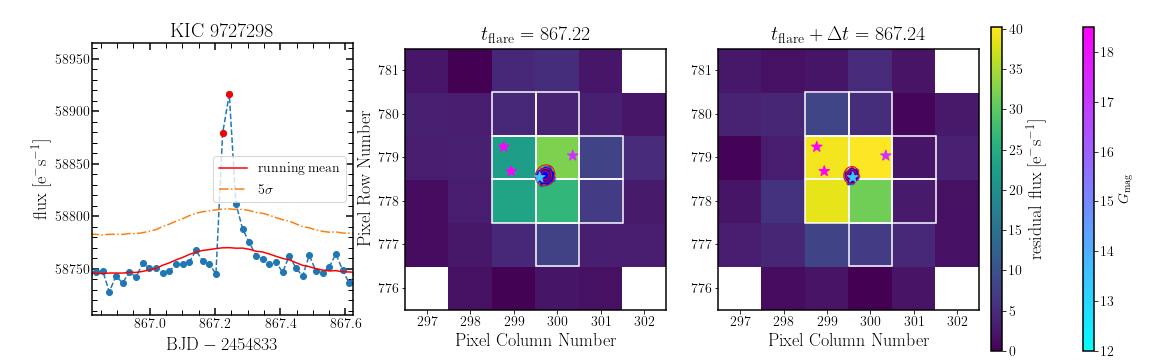}
             \caption{Two examples of flares associated with the target star. Symbols and colors are the same as in Fig.~\ref{fig:ellipces_on_the_sky_with_the_gaia_stars}.}
            \label{fig:flare_on_the_target_star}
 \end{figure*}

Figure~\ref{fig:flare_on_the_target_star}

 \newpage
\section{Flare on the background star.}\label{append:background}
\begin{figure*}[ht!]
    \centering
    \begin{subfigure}[t]{1.0\textwidth}
        \centering
                \includegraphics[width=\textwidth]{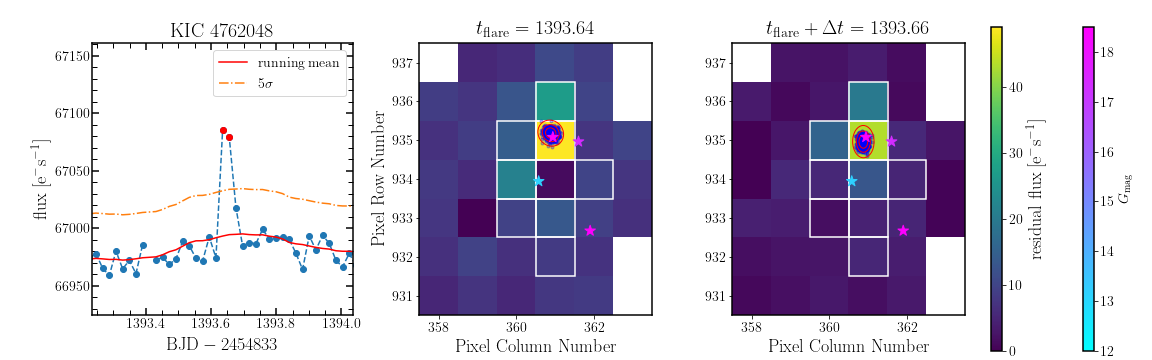}
        \includegraphics[width=\textwidth]{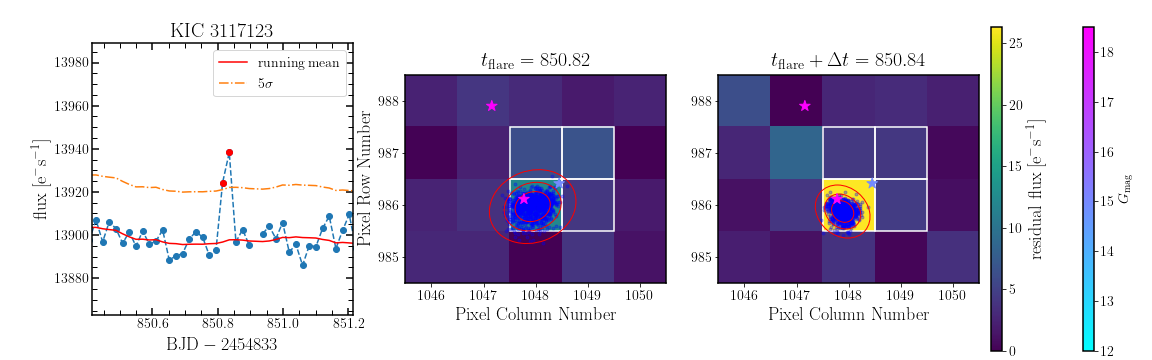}
        \includegraphics[width=\textwidth]{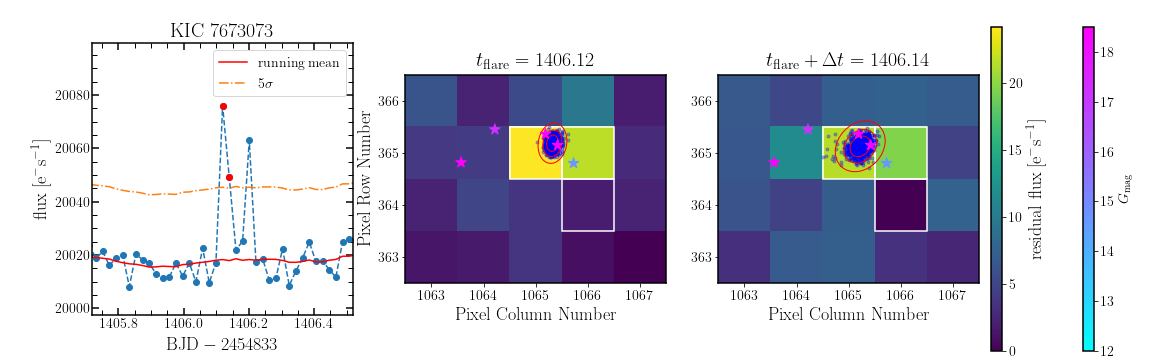}
    \end{subfigure}
    \caption{Three examples of flares happening on background stars. Symbols and colors are the same as in Fig.~\ref{fig:ellipces_on_the_sky_with_the_gaia_stars}.}
    \label{fig:flare_on_background_star}
\end{figure*}

Figure~\ref{fig:flare_on_background_star}

 \newpage
\section{Flares on target or background stars}
\begin{figure*}[ht!]
        \centering
                \includegraphics[width=\textwidth]{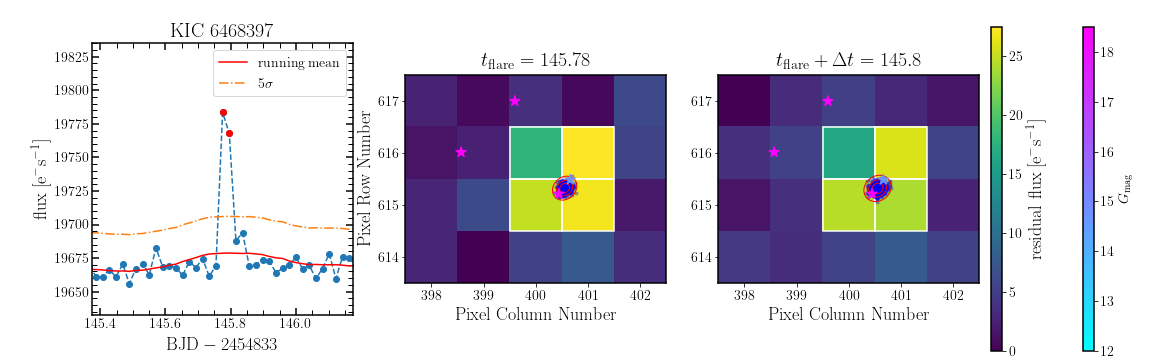}
                
                \includegraphics[width=\textwidth]{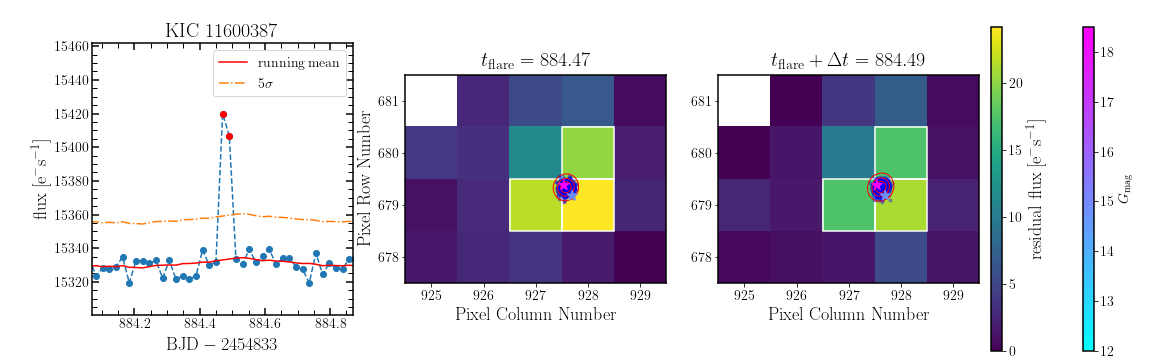}
             \caption{Two examples of events where the flare cannot be distinguished between the target and a background star. Symbols and colors are the same as in Fig.~\ref{fig:ellipces_on_the_sky_with_the_gaia_stars}.}
            \label{fig:multiple_stars_in_ellipce}
 \end{figure*}

Figure~\ref{fig:multiple_stars_in_ellipce}

\newpage
\section{Rejected cases}
\label{append:rejected_cases}
\begin{figure*}[ht!]
        \centering
            \begin{subfigure}[t]{1.0\textwidth}
        \centering
        \includegraphics[width=\textwidth]{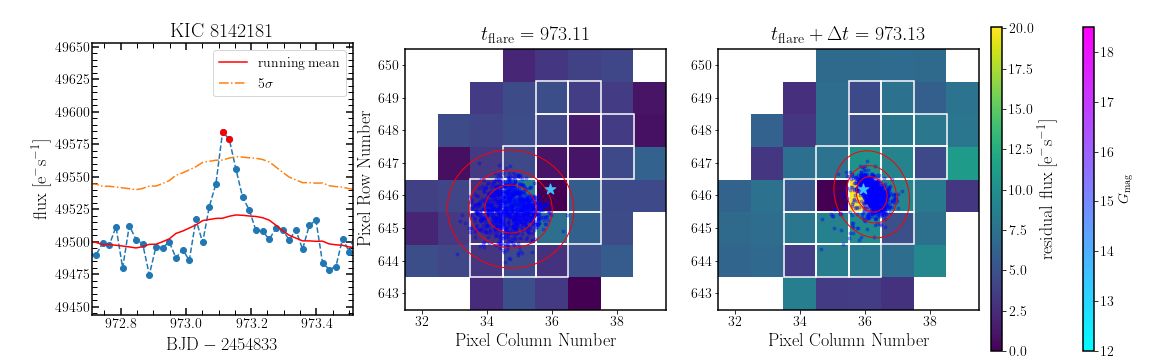}
        \includegraphics[width=\textwidth]{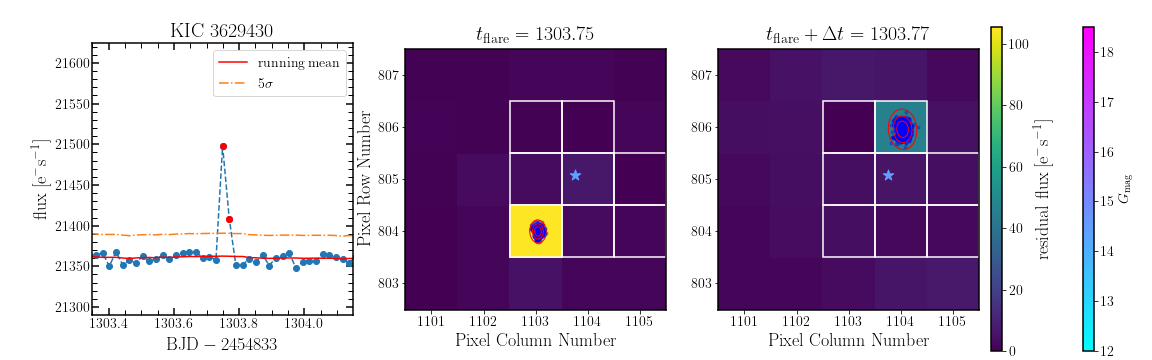}
        \includegraphics[width=\textwidth]{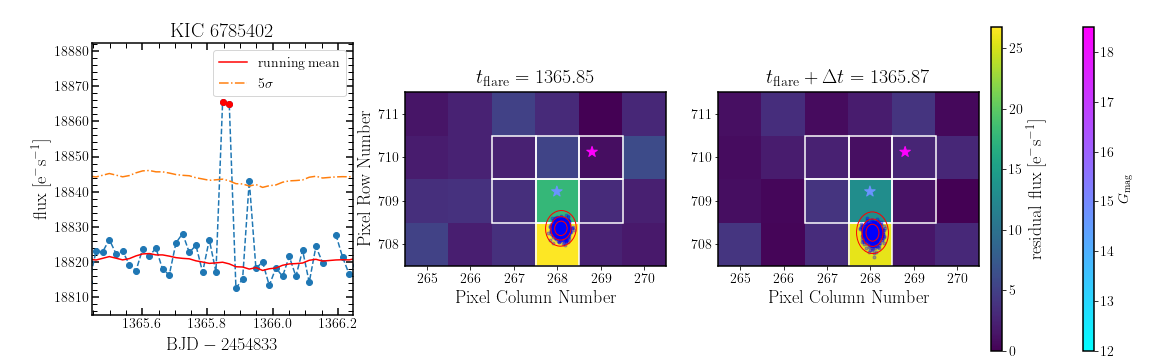}
                
    \end{subfigure}
             \caption{Rejected events. Top panels: transit of a solar system object across the image.
             Middle panels: the confidence ellipses of the two cadences ($t_\mathrm{flare}$ and  $t_\mathrm{flare} + \Delta t$) are located in different pixels.  
             Bottom panels: the confidence ellipses are at the same location for both cadences but are not associated with any star in the Gaia DR2 catalog. }
            \label{fig:ellipce_different_location}
 \end{figure*}

Figure~\ref{fig:ellipce_different_location}

\newpage
\section{Rejected case from Okamoto et al. (2021)}
\begin{figure*}[t!]
    \centering
    \begin{subfigure}[t]{1.0\textwidth}
        \centering
        \includegraphics[width=\textwidth]{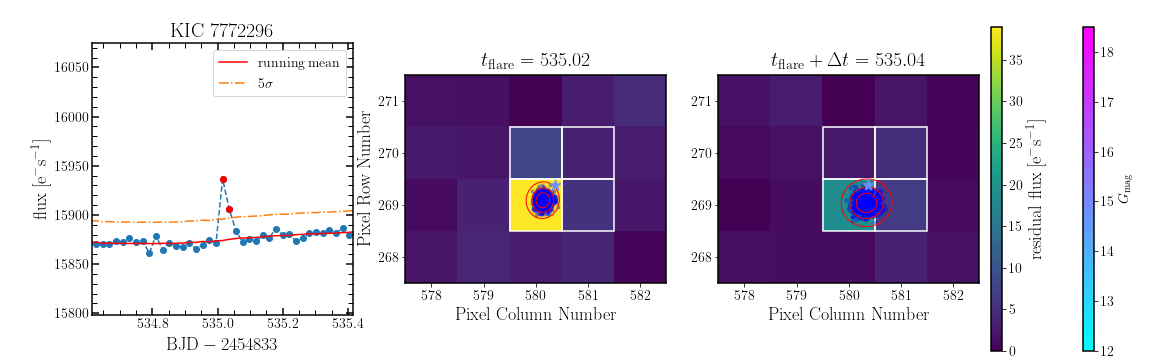}
        \includegraphics[width=\textwidth]{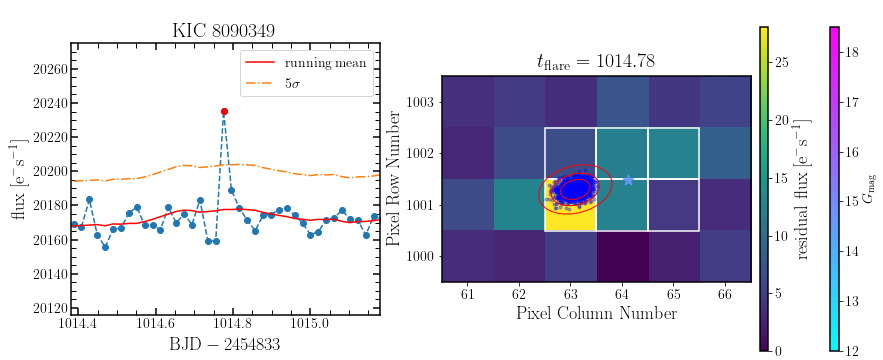}
        \includegraphics[width=\textwidth]{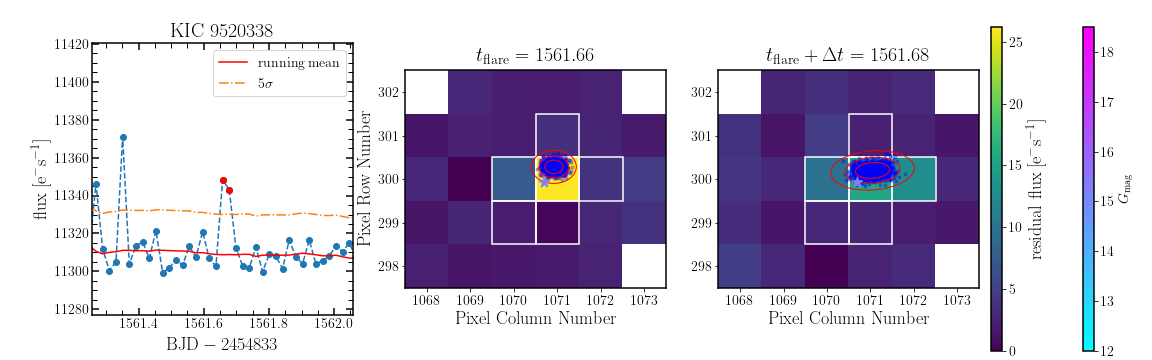}
        \includegraphics[width=\textwidth]{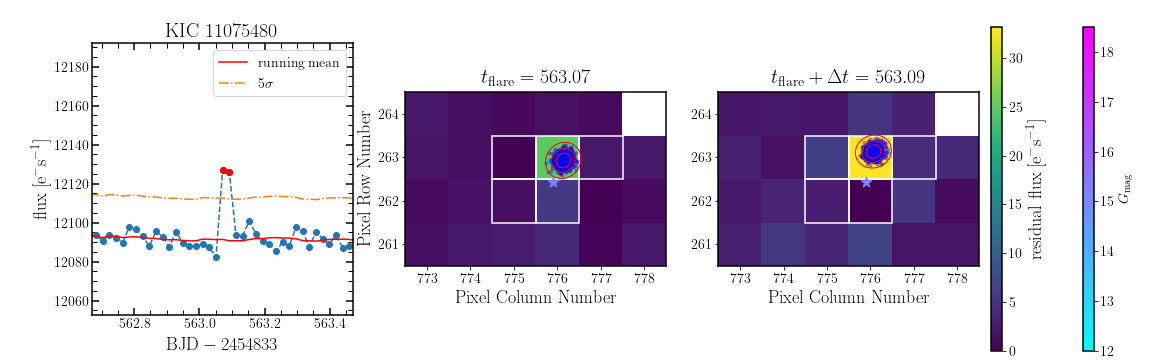}
    \end{subfigure}
\end{figure*}

\begin{figure*}[t!]\ContinuedFloat
    \centering
    \begin{subfigure}[t]{1.0\textwidth}
        \centering
        \includegraphics[width=\textwidth]{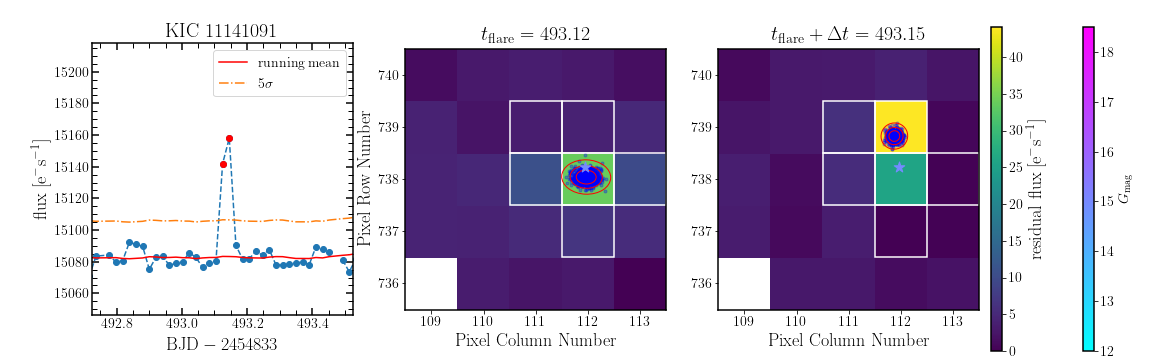}
        \includegraphics[width=\textwidth]{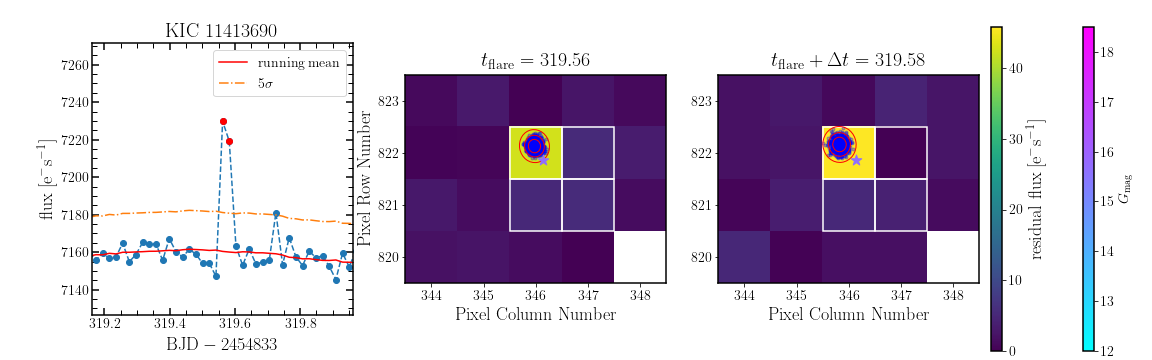}
        \includegraphics[width=\textwidth]{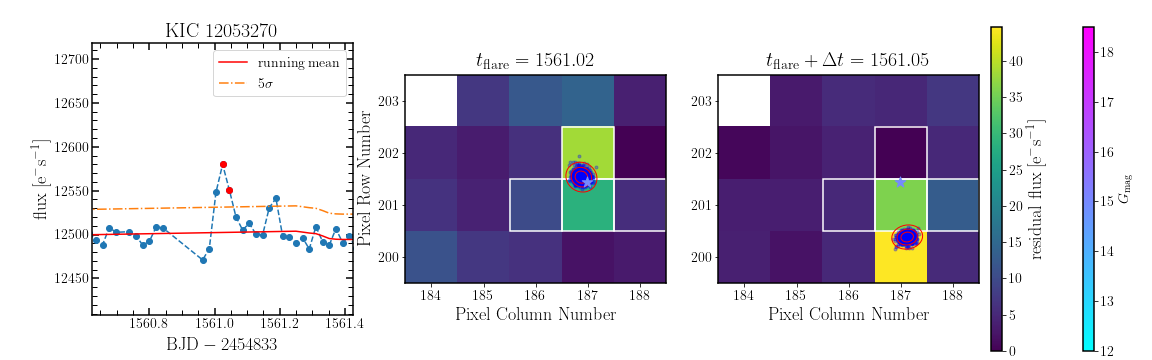}
    \end{subfigure}
    \caption{7 Sun-like stars with flare detections reported by \citet{Okamoto2021}, which have been rejected by our method.
    Left panel: light curve with the running mean (red line) and $5\sigma$ level (dashed-dotted orange line) used to find flare candidates. The red dots show the cadences ($t_\mathrm{flare}$ and  $t_\mathrm{flare} + \Delta t$) for which the PSF-fitting was performed. Middle and right panels: Residual images with the flare location and the 68\%, 95\%, and 99.9\% confidence ellipses. For events that have only one data point above the $5\sigma$ level on the light curve, we performed the PSF-fitting of the flare image only for that cadence.}
    \label{fig:okamoto_stars}
\end{figure*}
\end{appendix}

% WARNING
%-------------------------------------------------------------------
% Please note that we have included the references to the file aa.dem in
% order to compile it, but we ask you to:
%
% - use BibTeX with the regular commands:
%   \bibliographystyle{aa} % style aa.bst
%   \bibliography{Yourfile} % your references Yourfile.bib
%
% - join the .bib files when you upload your source files
%-------------------------------------------------------------------

\end{document}